\documentclass[10pt,journal,compsoc]{IEEEtran}

\AtBeginDocument{%
  }

\usepackage{amsmath}
\usepackage{booktabs}
\usepackage{cleveref}
\usepackage{graphicx}
\usepackage{float}
\usepackage{subfig}
\usepackage{multirow}
\usepackage{array}
\usepackage{enumitem}
\usepackage{amsthm}
\usepackage{tabularx}
\newtheorem{theorem}{Theorem}
\usepackage{color}
\newcommand{\rev}{\color{black}}

\crefname{section}{§}{§§}
\Crefname{section}{§}{§§}

\begin{document}


\title{FERN: Leveraging Graph Attention Networks for Failure Evaluation and Robust Network Design}


\author{Chenyi~Liu,
        Vaneet Aggarwal,
        Tian Lan,
        Nan Geng,
        Yuan Yang,
        Mingwei Xu,
        and Qing Li
 \IEEEcompsocitemizethanks{\IEEEcompsocthanksitem Chenyi Liu, Nan Geng, Yuan Yang and Mingwei Xu are with Department of Computer Science and Technology, Tsinghua University,
 and also with the
Beijing National Research Center for Information Science and Technology
(BNRist) (e-mail: liucheny19@mails.tsinghua.edu.cn;
nan\_geng@sina.com;  yangyuan\_thu@mail.tsinghua.edu.cn; xumw@tsinghua.edu.cn). Yuan Yang and Mingwei Xu are also with the Zhongguancun Laboratory.Nan Geng participated in this work during his PhD at Tsinghua University.
\IEEEcompsocthanksitem Vaneet Aggarwal is with Purdue University (e-mail: vaneet@purdue.edu).
\IEEEcompsocthanksitem Tian Lan is with George Washington University (e-mail: tlan@gwu.edu).
\IEEEcompsocthanksitem Qing Li are with the Peng Cheng Laboratory, (e-mail: liq@pcl.ac.cn).
}
}

\IEEEtitleabstractindextext{
 \begin{abstract}
    Robust network design, which aims to guarantee network availability under various failure scenarios while optimizing performance/cost objectives, has received significant attention. Existing approaches often rely on model-based mixed-integer optimization that is hard to scale or employ deep learning to solve specific engineering problems yet with limited generalizability. In this paper, we show that failure evaluation provides a common kernel to improve the tractability and scalability of existing solutions. By providing a neural network function approximation of this common kernel using graph attention networks, we develop a unified learning-based framework, FERN, for scalable Failure Evaluation and Robust Network design. FERN represents rich problem inputs as a graph and captures both local and global views by attentively performing feature extraction from the graph. It enables a broad range of robust network design problems, including robust network validation, network upgrade optimization, and fault-tolerant traffic engineering that are discussed in this paper, to be recasted with respect to the common kernel and thus computed efficiently using neural networks and over a small set of critical failure scenarios. Extensive experiments on real-world network topologies show that FERN can efficiently and accurately identify key failure scenarios for both OSPF and optimal routing scheme, and generalizes well to different topologies and input traffic patterns. It can speed up multiple robust network design problems by more than 80x, 200x, 10x, respectively with negligible performance gap.
 \end{abstract}
}
\maketitle

\section{Introduction}
\label{sec:intro}

As the scale and complexity of modern networks continue to increase rapidly, the occurrence of failures has become a common and frequent event in both wide area networks (WANs) \cite{markopoulou2008characterization,turner2010california,zheng2019sentinel,govindan2016evolve,hong2018b4,krishnaswamy2022decentralized} and data center networks (DCNs) \cite{gill2011understanding}. 
Recently, increasing research efforts have considered tackling the problem from different aspects.
Some studies consider the network planning stage  \cite{zhu2021network,ahuja2021capacity} to minimize performance variations with potential failures.
Some study quantifies network performance under the worst failure case \cite{chang2017robust}.
There are also studies focusing on the traffic engineering perspective  \cite{wang2010r3, liu2014traffic, bogle2019teavar, chang2019lancet, jiang2020pcf}, which make fault-tolerant management decisions for operating networks.
We jointly name these problems, including but not limited to robust network planning, robust network validation, and robust traffic engineering, as \emph{robust network design} problems.
Most existing approaches for robust network design rely on model-based mixed-integer optimization and are hard to scale to large networks.

We observe that a cornerstone in many robust network design problems is to ensure network availability under different failure scenarios for potential solutions.
Unfortunately, the number of failure scenarios grows exponentially with network size due to the combinatorial nature of failures.
Consequently, ensuring the network availability under millions of failure combinations can quickly become computationally prohibitive, constituting a key challenge in robust network design.
For example, solving a fault-tolerant traffic engineering problem for a 100-node network via linear programming (LP) could take several weeks \cite{abuzaid2021contracting}.
Such overhead is unacceptable in dynamic environments where network topology and traffic characteristics are time-varying.
Most existing approaches simply enumerate the failure scenarios \cite{zhu2021network, bogle2019teavar}, and each failure scenario should be considered separately, e.g., as a constraint in the optimization model.
Some approaches relax the integer variables that indicate failures into continuous variables, and then solve the optimization problem with approximation methods \cite{chang2017robust,chang2019lancet}. 
However, these approaches still incur a large computation overhead or could lead to severe performance degradation \cite{chang2019lancet,jiang2020pcf}.
Furthermore, most existing approaches propose to transform their optimization models for a solution that can balance optimality and tractability, but the transformation methods are only suitable for specific problems.
A generalized approach to enhance computation efficiency is strongly desired for robust network design problems.

In this paper, we identify failure evaluation as a common kernel to address the cornerstone in robust network design. 
This paper develops a unified learning-based framework, FERN, for scalable Failure Evaluation and Robust Network design.
FERN leverages graph attention networks (GATs) \cite{brody2021attentive} for scalable failure impact evaluation.
In particular, FERN takes network topology, traffic demand, route decisions, and target failure scenarios as inputs, and generates failure impact predictions (formally defined as the normalized maximum link utilization under given failures) in a one-shot computation.
Through such neural network function approximation, FERN provides a common kernel for formulating and solving many robust network design problems, enabling a broad range of robust network design problems to be recasted with the common kernel and thus computed efficiently using neural networks.

Machine Learning (ML) has achieved great success in many networking problems such as congestion prediction \cite{poularakis2021generalizable}, network planning \cite{zhu2021network}, and traffic engineering (TE) \cite{xu2018experience}.
{\rev
However, most existing solutions are tailored to specific engineering problems, making it challenging to apply them to different design objectives or formulations. For instance, a model designed for traffic engineering, which focuses on determining link weights for steering traffic demand, cannot be directly applied to network planning, which involves designing both short-term and long-term network topologies. These two problems have distinct inputs, outputs, and underlying principles. 
}
In contrast, we advocate a new generalizable approach harnessing ML for robust network design.
We demonstrate the generalizability of FERN through three use cases: robust network validation, network upgrade optimization, and fault-tolerant traffic engineering.
By recasting robust network design problems using the approximated kernel and thus identifying a small number of critical failure scenarios that significantly impact the optimizations (while pruning the inconsequential ones), FERN allows these problems to be computed with great efficiency.

Deriving a neural network function approximation for accurately evaluating the impact of different failure scenarios is not an easy task.
Complex problem information, including network topology, traffic demand, routing decisions, and failure scenarios, as well as their relationships, must be effectively captured and embedded into feature representations.
To address the issue, we first provide several key insights into the failure impact, which are generalized under different routing schemes.
Based on these common insights, we propose a novel GAT-based architecture to predict failure impact, in which the network topology, traffic demand, route, target failure scenarios, and other key features are embedded and represented as a graph.
Then, we design a GAT with both local and global views to attentively perform feature extraction from the graph and to predict the outcome of different link failures, enabling generalizability and computational efficiency.
We train FERN on a large dataset containing network topologies with different scales to obtain a highly generalizable pre-train model.
For each topology in the training set, we generate a few failure scenarios and label them with failure impact and important weights by simulating the rerouting process.
We construct a novel loss function using the labels for failure impact and important weights.
As a result, FERN is able to distinguish critical failure scenarios from insignificant ones that can be pruned.
To further improve the model accuracy in specific problems, FERN supports a second training phase with negligible time cost.


We conduct extensive experiments to evaluate FERN with real-world network topologies at different scales.
Evaluation results show that FERN could predict the failure impact accurately with satisfactory generalizability and significant computational overhead reduction.
Further experiments validate that with the help of our proposed failure evaluation and critical failure detection algorithm, the time reduction of robust network design problems could be up to 200x (i.e., two orders of magnitude) compared to existing optimization approaches while obtaining the near-optimal solution.


The main contributions of our work are as follows:
\begin{itemize}
\item We show that efficient failure evaluation could be a common kernel in robust network design problems to reduce the number of failure combinations that grow exponentially with network size, addressing the scalability issues in existing robust network design approaches.
\item To this end, we analyze common insights of the failure impact under different routing schemes and leverage a novel GAT-based architecture to propose a neural network function approximation solving the kernel problem, failure evaluation, in a one-shot calculation.
\item FERN is able to support a broad range of robust network design problems. We demonstrate FERN’s generalizability using three important use cases: robust network validation, network upgrade optimization, and fault-tolerant traffic engineering. We present and prove performance guarantees of applying FERN to solve such robust network design problems.
\item The proposed FERN framework and the three use cases are evaluated on real-world network topologies, matching the accuracy of state-of-the-art baselines but delivering orders of magnitude speedup in time-to-solution.

\item We further analyze the potential of using FERN to address failure evaluation and robust network design problems for real-world routing schemes that may not be formalized with LP models easily, e.g., OSPF. 
\end{itemize}

Employing machine learning for network optimization has been an active area of research that received significant recent attention \cite{zhu2021network, xu2018experience, geng2020multi, bernardez2021machine}. Nevertheless, our approach in this paper is unique. It illuminates a new direction toward integrating Artificial Intelligence (AI) and networking – by identifying common cornerstone tasks in general networking problems and then building their neural network function approximations as reusable kernels. It could be the first step toward a new class of algorithms for network optimization.

\section{Background and Motivation}

\begin{table*}[ht]\footnotesize 
    \caption{Computational overhead and time complexity of three classic robust network design problems (discussed in detail in Section \cref{sec:fault-tolerant-network-design}) under 2 simultaneous link failures\protect\footnotemark. Letter `F' indicates that optimization solver exceeded the memory limit of the sever and failed to yield a solution in our experiment.} 
    \label{tab:motivation-problemscale}
    \centering
    \scalebox{1}{
    \begin{tabular}{ccccccc}
        \toprule
        \textbf{Network} & \textbf{Nodes} & \textbf{Edges} & \textbf{Failure scenarios} & \textbf{Robust validation} & \textbf{Network upgrade} & \textbf{Fault-tolerant TE}\\
        \midrule
        Abilene & 11 & 14 & 94 & 5s & 4s & 0.5s\\
        Geant2010 & 31  & 50 & 1255 & 2722s & F & 613s \\
        DialtelecomCz & 106 & 119 & 6786 & 3032940s & F & F\\
        \bottomrule
        \end{tabular}
    }
    \vspace{-3mm}
    \end{table*}

Recent works \cite{abuzaid2021contracting, krishnaswamy2022decentralized} by Microsoft show that modern wide-area-network (WAN) are becoming larger and larger with thousands of routing nodes and are capable of taking larger-scale traffic flow. 
Moreover, the topology size keeps growing at a rate of 20\% per year \cite{zhu2021network} with no end in sight. 
For such large-scale networks, it is an important yet challenging task to guarantee network availability for diverse failure scenarios in the data plane, control plane, and management plane. 
The evolving network architecture and traffic demands make the problem even worse.  

In this section, we first illustrate the limitations of existing works for robust network design in the face of large-scale network topologies.
Then we show the motivation for designing GAT-based FERN to fast predict failure impact and detect critical failure scenarios.

\subsection{Limitations in Existing Approaches}

Network availability under failure scenarios has always been an important part of network design~\cite{chang2017robust,chang2019lancet,jiang2020pcf,liu2014traffic,bogle2019teavar,wang2010r3,zhu2021network}. 
Examples include three popular fault-tolerant network design problems as follows
\begin{itemize}[leftmargin=15pt,topsep=5pt]
    \item \emph{Robust network validation}: Finding the worst performance (e.g., max link utilization) of a network design under a set of failure scenarios.
    \item \emph{Network upgrade optimization}: Minimize the cost of added capacities to the links in the network to satisfy congestion-free constraints under a set of failure scenarios. 
    \item \emph{Fault-tolerant traffic engineering}: Finding the route and reroute decision under a set of failure scenarios to balance the network utilization and availability.
\end{itemize}

Existing works for robust network design usually build an optimization model to optimize the network utilities under various failure scenarios. For instance, robust network validation solves a max-min optimization problem based on MCF models for all the possible failure scenarios \cite{chang2017robust}, while network upgrade optimization and fault-tolerant traffic engineering solve LP or ILP  problems with congestion-free constraints for each failure scenario \cite{zhu2021network,wang2010r3}.   We note that a common core of these robust network design problems is to enumerate all possible failure scenarios, model the failure impact, and validate or constrain the worst performance under the possible failure scenarios. More details of the three problems will be provided in  \cref{sec:fault-tolerant-network-design}.

The common core poses a challenge to the scalability of existing algorithms with the rapid growth of possible failure combinations in current networks. For instance, many recent proposals studied ensuring the network performance under $f$ simultaneous link failures \cite{wang2010r3, chang2019lancet}. 
We show the computational overhead to solve the three robust network design problems on a server with 104 core CPU and 256 GB memory using gurobi \cite{gurobi}, a state-of-the-art optimizer, for topologies with different scales in Table. \ref{tab:motivation-problemscale}.
We note that only considering $f$ simultaneous link failure scenarios would bring $O(|E|^f)$ failure scenarios under consideration, causing a super-linear growth in the LP/ILP problems scale with the network scale increase.
Moreover, the solution time to such LP/ILP problems also increases super-linearly with the number of decision variables and constraints \cite{abuzaid2021contracting}.
The two factors mentioned above jointly make robust network design problems hard to solve with increasingly large and complex network topologies and failure scenarios in the real world. 
Furthermore, both the combination of failures and the routing scheme could be much more complicated in practice. The complicated failures and routing schemes make it even harder to model the impact of all the possible failures and guarantee the availability under failures in a typical LP/ILP problem.

\footnotetext{We remove failure scenarios which split the network into two parts to ensure that there is a feasible solution to the problem.}

\subsection{Motivation of FERN}

\begin{figure}[htbp]
\centering
    \subfloat[\footnotesize OSPF routing.]{
        \begin{minipage}[b]{0.23\textwidth}
        \centering
        \includegraphics[width=1\textwidth]{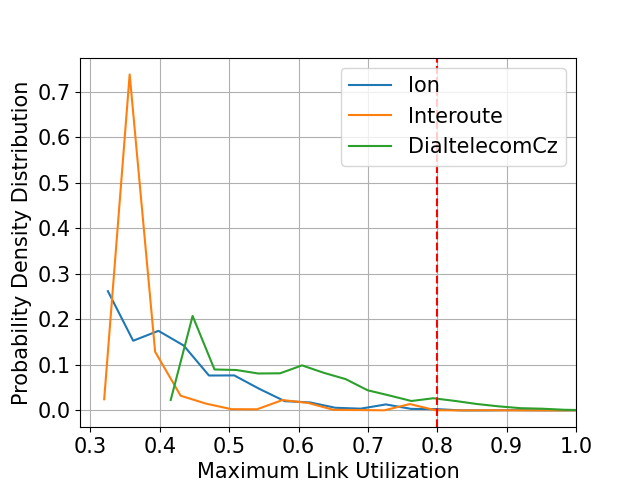}
        \end{minipage}
    }
     \subfloat[\footnotesize Optimal (MCF) routing.]{
        \begin{minipage}[b]{0.23\textwidth}
        \centering
        \includegraphics[width=1\textwidth]{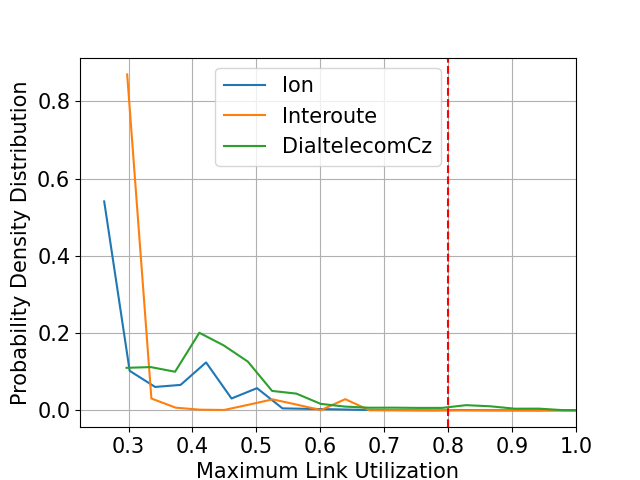}
        \end{minipage}
    }
    \captionsetup{font={footnotesize}}
    \caption{Distributions of MLU increase on large-scale topology for OSPF and optimal (MCF) routing schemes under 2 simultaneous link failures. MLU under failure scenarios are normalized by the MLU under the worst-case failure scenario.}
    \label{Fig:failure-case-Ion}
\end{figure}

As mentioned, the combinatorial nature of failure scenarios poses a core challenge to robust network design.
As shown in Table \ref{tab:motivation-problemscale}, thousands of failure scenarios formed by the combination of two links make robust network design problems large and hard to solve for large-scale network topology.
To this end, we observe that for large-scale network topologies, only a small subset of failure scenarios are critical.
Fig.~\ref{Fig:failure-case-Ion} shows the distribution of failure impact (i.e., maximum link utilization (MLU) increase) under OSPF \footnote{We set the reciprocal of the link bandwidth as the weight of OSPF.} and optimal (MCF) routing scheme for three real-world large scale network topologies with more than 100 nodes. 
It turns out that only 0.19\%, 0.03\%, and 3.43\% failure scenarios on Ion, Interoute, and DialtelecomCz, respectively, under the optimal routing scheme, cause significant impact (i.e., more than 80\% of worst-case failure impact) to the network availability.
The results for the OSPF routing scheme are 0.10\%, 0.06\%, and 5.82\%, respectively.
We note that the ratio of critical failure scenarios causing significant impact is small on large topologies. 
It implies that by providing an approximation of the failure impact evaluation function, we could prune many unimportant failure scenarios and focus only on a small subset of critical failure scenarios with large failure impact in robust network optimization.

Let $x$ be a failure scenario caused by several link failures.
When such a common failure scenario occurs on bottleneck links, a large amount of traffic may be blocked and need to be rerouted, which will seriously affect network performance and cause network congestion.
This motivates us to define a general failure impact evaluation function as
\begin{equation}
    impact(x)=F(x, G, D, r).
    \label{eq:impact-def}
\end{equation}
In other words, the impact of a failure -- defined in this paper as the increase of maximum link utilization (MLU) under given failure scenario $x$, is determined by network topology $G$, traffic demand $D$, and routing decision $r$.
In this paper, we will show that if an efficient approximation of $F(x, G, D, r)$ can be obtained, e.g., through the use of neural networks, many robust network design problems could be recast and computed with ease relying on the approximation.
For instance, we can obtain the worst-case network performance under a set of failure scenarios as long as we can accurately predict the impact of failures.
We further discuss how we can recast the robust network design problems using the failure impact function $F$ in \cref{sec:fault-tolerant-network-design}.

Unfortunately, modeling the failure impact is quite a challenging task. For instance, in a theoretically optimal setting
we need to solve an MCF problem for each failure scenario to simulate the failure impact. Moreover, practical network
failures may be caused by a combination of failures in the data plane, control plane, and management plane, which makes it even more difficult to model the impact of a failure combination. 
The complexity and the huge number of possible failure combinations make it quite difficult to figure out the critical failure scenarios causing significant impacts on the network.

Motivated by these key observations, we explore the potential of machine learning in resolving the common core of failure evaluation for robust network design.
It provides a neural network approximation of the failure impact function $F_{\theta}(x, G, D, r)$ and enables us to quickly identify a small set of critical failures that are most relevant to robust network design. 
Compared to traditional approaches, which simulate the network's behavior in each failure scenario (e.g., solving an MCF problem for optimal reroute), deep learning-based approaches allow us to efficiently assess the impact of failures even with incomplete or erroneous network information. 
Recently, graph neural networks (GNNs) have been applied in network planning \cite{zhu2021network}, traffic engineering \cite{bernardez2021machine} and congestion prediction \cite{poularakis2021generalizable} to extract the network features represented as graph.
The evaluations show that GNNs can achieve good performance in such tasks and has the potential to generalize to different network topology.
GAT \cite{brody2021attentive} is a state-of-the-art model of GNN which shows good capability and generalizability in different application scenarios.
In this work, we propose a GAT-based mechanism to predict the link failure impact and select the critical failure scenarios enhancing robust network design.

\section{Overview}
\label{sec:overview}

\begin{figure*}[htbp]
    \centering
    \includegraphics[width=0.8\linewidth]{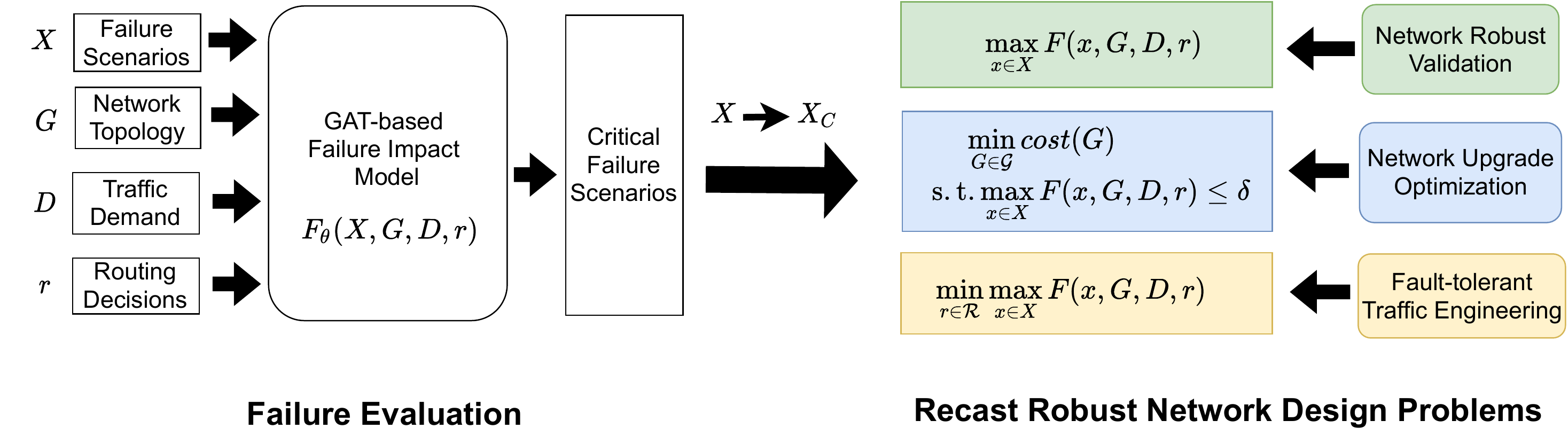}
    \caption{Overall structure of FERN.}
    \label{Fig:overview}
\end{figure*}

In this paper, we aim to resolve the robust network design problems in two steps.
We show the general perspective of our approach in Fig. \ref{Fig:overview}.
First, we design a GAT-based function approximation $F_\theta$ to predict the impact of target link failure scenarios and figure out critical failure scenarios.
The GAT-based function $F_\theta$ takes target failure scenarios, network topology,  traffic demand, and routing decision as input, and outputs predicted impact of the target failure scenarios in a one-shot inference.
With the predicted results, we select a small subset of critical failure scenarios $X_C$ from the full failure set $X$.  
Such a failure evaluation algorithm should have the characteristics as described in the following:
\begin{itemize}[leftmargin=15pt,topsep=15pt,partopsep=0pt]
    \item \emph{High computational efficiency}: The algorithm should have low time cost and memory use, and keep a low overhead increase when the topology scale increases. 
    \item \emph{High accuracy}: The algorithm should predict the impact of target failure scenarios accurately, especially for the potentially critical failure cases.
    \item \emph{Good generalization}: The algorithm should have good generalization to unseen network topologies, traffic demands, and other possible application scenarios.
\end{itemize}
Note that compared to existing works that apply deep learning for network optimization, generalization could be a critical point for our work. In this way, we could reduce the computational overhead under dynamically changing network environments.

We recast the robust network design problems using the failure impact function $F$ and then enhance the tractability of the recast problems with the selected critical failure scenarios.
In \cref{sec:fault-tolerant-network-design}, we show how to apply FERN in three typical robust network design problems to resolve the problem efficiently.
In the robust network validation problem, we show that we could obtain the worst-case network performance only by validating the critical failure scenarios.
In the network upgrade optimization problem, we show that only adding congestion constraints to the worst failure scenarios could obtain the exact optimal solution to the original problem.
In the fault-tolerant traffic engineering problem, we show that designing the routing strategy for critical failure scenarios with an additional load balancing objective can bring good network availability at a small cost of computational overhead.
In addition to the three use cases above, we believe FERN could be applied to many existing robust network design problems.

\section{Proposed Fern Framework}
\label{sec:methodology}
In this section, we present a graph-based deep learning mechanism to efficiently predict the failure impact and detect the critical failure scenarios.
First, we introduce the definition and key insights of failure impact.
Second, we propose a GAT-based failure impact prediction model based on the insights.
Third, we design a loss function to train the impact prediction model efficiently.
Finally, we briefly explain how we train the model and apply it to detect critical failure scenarios.

\subsection{Impact of Failure Scenarios}
\label{sec:impact-insights}

Maximum link utilization (MLU) is a popular performance metric in existing works for network planning and traffic engineering to measure the network congestion level \cite{chang2017robust}. 
In this paper, we use MLU increase under a failure scenario to measure the failure impact. 
In particular, for failure scenario $x$, we apply a deep learning mechanism $\theta$ to approximate the failure impact in FERN, defined as
\begin{equation}
    F_\theta(x,G, D, r) = \frac{MLU(x, G,D,r)}{MLU(\phi, G, D, r)}, ~ x \in X. \label{eq:impact-def-FERN}
\end{equation}
where $MLU(x,G,D,r)$ denotes MLU of network $G$ under traffic demand $D$, routing decision $r$ and failure scenario $x$, and $MLU(\phi,G,D,r)$ denotes MLU under non-failure scenario. 
MLU increase indicates the degree of congestion increase in the network under failure scenarios, and is independent of the input traffic level since it is normalized by MLU in non-failure scenario.
We note that MLU increase enables a unified measure of failure impact across different topologies and traffic matrices, and makes the range of failure impact consistent.

Ideally, we could obtain the failure impact by approximating the routing and rerouting process (e.g., solving multi-commodity flow (MCF) problems or shortest path routing problems) under non-failure scenario and all the possible failure scenarios.
However, designing a deep learning mechanism to solve such complex optimization problems is a non-trivial task.
To this end, we analyze the rerouting process of optimal routing scheme, and give some common insights of failure impact, which guide us to design an efficient and generic model approximation for failure evaluation.


\begin{figure}[htbp]
\centering
    \subfloat[\footnotesize Overall relative error.]{
        \begin{minipage}[b]{0.23\textwidth}
        \centering
        \includegraphics[width=\textwidth]{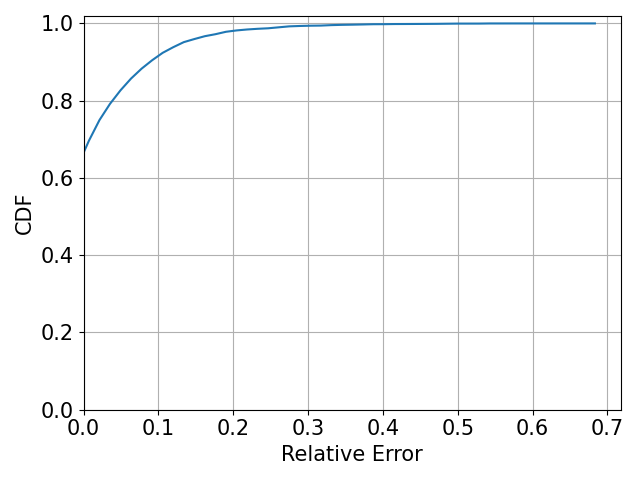}
        \end{minipage}
    }
     \subfloat[\footnotesize Failure impact on Tw topology.]{
        \begin{minipage}[b]{0.23\textwidth}
        \centering
        \includegraphics[width=\textwidth]{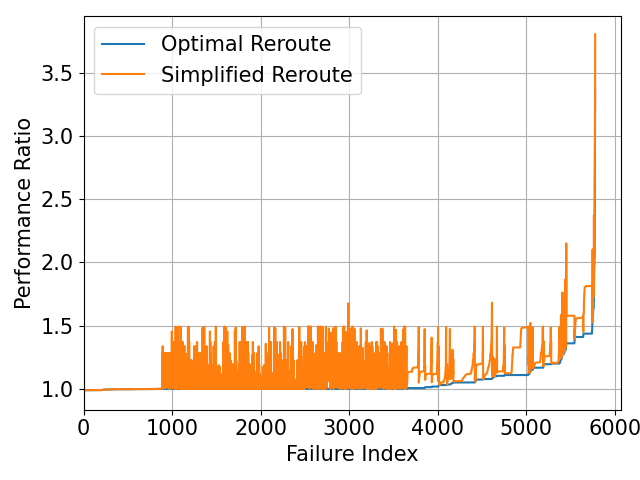}
        \end{minipage}
    }
    \captionsetup{font={footnotesize}}
    \caption{ Comparison of failure impact between optimal reroute and simplified reroute.}
    \label{Fig:reroute-analysis}
\end{figure}

In order to understand the impact of link failures intuitively, we would like to introduce a simplified reroute problem as an approximation to the original optimal (MCF) reroute problem.
In particular, when a failure scenario occurs, the simplified problem only reroutes the flows that pass through the failed links.
We compare the impact of 2 simultaneous link failures under the original optimal reroute and the simplified reroute over more than 100 real world topologies in topology zoo.  
The results are shown in Fig. \ref{Fig:reroute-analysis} (a).  
We find that the difference in failure impact between the optimal and simplified reroute is less than 10\% for more than 90\% failure scenarios. 
We further analyze cases for the topologies with large relative errors and use the Tw topology as an example to show the results in Fig. \ref{Fig:reroute-analysis} (b).
We find that the failure scenarios that have large impact under the optimal reroute still have a high impact under the simplified reroute, which indicates the rationale for selecting the critical failure scenarios with the simplified reroute problem.

With the aspect of the approximate simplified reroute problem, we show some key insights of failure impact under the optimal routing scheme.

\textbf{Concentrated traffic}: Although MCF problem does not limit the number of paths for each flow, the optimal routing decision usually concentrate the traffic of a flow demand into 1 or 2 paths.  Thus we can represent the optimal route in the form of paths and corresponding flow assignments. 

\textbf{Critical indicators}: The impact of a link failure scenario is decided by input traffic and network topology. 
In this work, we find that the network topology is the primary factor in our experiment, while the input traffic matrix also influences the results. 
Thus, we embed the traffic demand into the network topology.
Specifically, we find that the link utilization and traffic volume passing through the link under non-failure optimal routing decision are two important features indicating a critical link failure. 
A link with higher link utilization and traversed traffic volume under the optimal routing decision may cause a greater impact on the network. 

\begin{figure*}[htbp]
    \centering
    \includegraphics[width=0.9\linewidth]{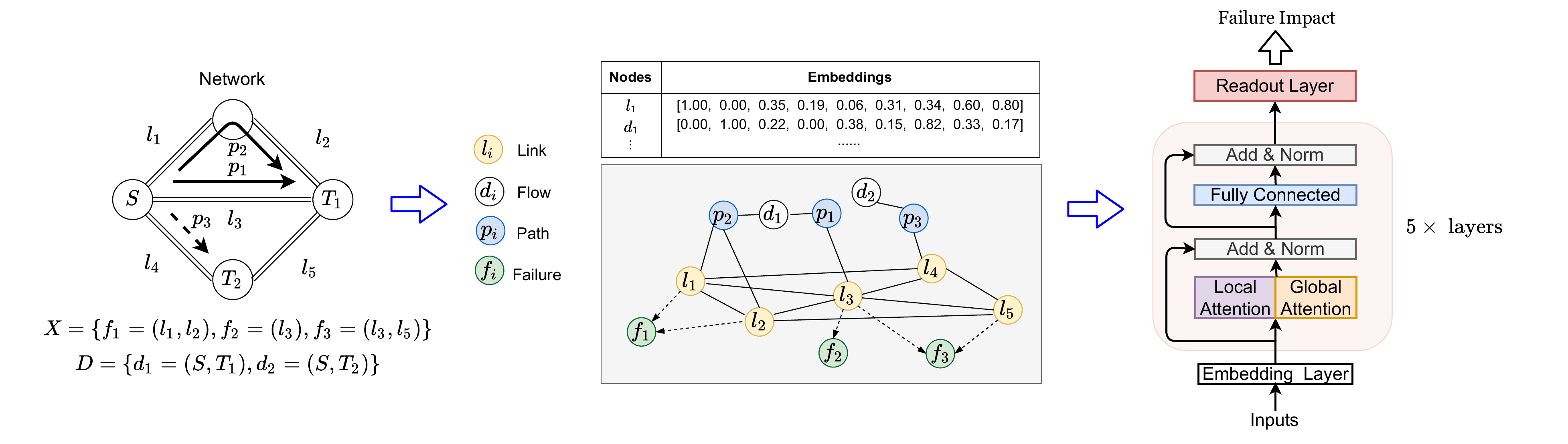}
    \caption{Working process of GAT-based failure impact prediction model.}
    \label{Fig:structure-GAT}
\end{figure*}

\textbf{Locality}: 
Network topology and traffic demand in current real world WAN are clustered locally \cite{abuzaid2021contracting}. 
When a failure occurs, finding a reroute locally to walk around the failed link could be a good choice in many situations since a local reroute can save the overall network capacity and cause a smaller impact on the whole network. 
Thus, the topology structure near a failed link is important to predict the impact of the failed link. 

\textbf{Long flow reroute}: Although many traffic flows can be near optimally rerouted locally, some large and long flows may need a long path (e.g., when the failed link is a bottleneck link). 
For such cases, we need a global understanding of the overall network topology and traffic route to estimate the impact of rerouting the flow by a long alternative path to walk around the congested area.  

\textbf{Monotonous failure influence combination}: We find that the impact of multi-link failure scenarios usually has positive correlations with the corresponding single-link failure scenarios. 
However, the exact correlation between the impact of multi-link failure and single-link failure could not be obtained with a simple linear combination. Further, the real correlation of the impact depends on the exact network topology and input traffic features.

We elaborate on the insights of failure impact with the optimal route and reroute based on the MCF problem above.
In practice, WAN may only allow for a small number of parallel paths per flow, such as in Equal-Cost Multi-Path (ECMP) and K-shortest path routing, where the optimal route and reroute are not feasible.
However, the MCF model is still widely used in practical network planning scenarios even with the non-optimal routing issue \cite{ahuja2021capacity}. 
Furthermore, we note that these key insights apply to many other commonly used routing schemes like shortest-path-based routing protocols.



\subsection{Failure Evaluation Model Design}
\label{sec:model-design}
{\rev 

The failure evaluation model takes network topology, input traffic, target failure scenarios, route, and reroute as inputs and predicts the impact of link failures using Eq. \eqref{eq:impact-def-FERN}. Since obtaining rerouting decisions for all target failure scenarios is challenging, we incorporate the rerouting policy into the Graph Attention Network (GAT)-based impact evaluation model. This means the trained model predicts the impact of target failure scenarios under a specific rerouting strategy (e.g., MCF or OSPF) given the network topology, input traffic, and original routing decision.
However, directly inputting the network topology, demand matrix, route, and target failure scenarios into a deep neural network doesn't yield an accurate and general model. To address this, we propose a novel deep learning architecture based on Graph Attention Network (GAT) in this section, guided by insights from \cref{sec:impact-insights}.


GAT is a type of graph neural network that leverages attention mechanisms to process graph data. Specifically, GATs embed input states into a directional graph, where each node in the graph is associated with a feature vector that represents information such as link capacity, link utilization, flow traffic demand, and node type. The key idea behind GATs is neighborhood aggregation, which involves computing a linear combination of the feature vectors of a central node and its neighbors for each node in the graph. The GAT accomplishes this by using an attention mechanism to learn weights that determine the importance of each neighbor's features for the central node.

GATs' ability to capture the influence of each node's neighbors in the graph makes it suitable for embedding and representing network features in failure evaluation. It enables identification of key information of local network structure and traffic assignments (represented as graph data), e.g., how neighboring links are affected by a failed link in the network.


In this work, we embed the correlation between traffic flows, routing paths, links and target failures in the network into a graph. Then, we leverage GAT and its attention mechanism to weigh the influence from neighboring nodes in the input graph and to represent the key information for inferring the failure impact in each failure scenario.
We also introduce an attention mechanism to weigh the failure influence from a broader aspect. 
In addition to aggregating information from neighbors in standard GAT, a global attention mechanism enables each link to further aggregates information and weigh the failure influence from all the other links in the network. Thus, the resulting model could better evaluate the impact of failures by combining global and local information.
}

\textbf{Input graph}: The procedure for building an input graph is shown in Fig. \ref{Fig:structure-GAT}. 
{\rev 
To construct a suitable input graph for our GAT-based algorithm, we firstly transform the original network topology to build up a basic input graph. In the transformed graph, each link in the original topology is transformed into a node, and a link is added between two nodes in the transformed graph if their corresponding links share a common endpoint in the original topology. 
The converted graph focuses on the link states and the connections between links in the original topology, making it easier for the GAT-based model to extract key information about the failure impact. 
}
Secondly, we model traffic demand and the routing decision of the flow demand under a non-link-failure scenario on the transformed graph. 
For each flow demand, we build up edges on the input graph between the flow node and the corresponding path nodes,  representing the routing paths of a flow.
And for each routing path,  we build up edges between the path node and the corresponding link nodes . 
For the last step, we model the target failure scenarios.
For each failure scenario, we build up the edges from the nodes of corresponding failed links to the failure node. 
The failure nodes aggregate link states for the final failure impact prediction.  
We also note that such an input graph can generalized to many routing schemes link optimal MCF, OSPF and ECMP. 

\textbf{Input state}: For the four types of nodes in the input graph, we design the initial state craftily to embed the link attributes, input traffic demand, and routing decision into the input graph. 
With the insights in \cref{sec:impact-insights}, we use link utilization and traffic volume under original routing decision, and link capacity as the initial state to represent the importance of the link. 
For the flow node, we use the flow demand as the initial state. 
For the path node, we use the traffic assignment, i.e., traffic split ratio, on the path as the initial state.
Moreover, to distinguish the node type, we set extra units in the state to indicate the node type. 
We normalize the link utilization, link capacity, and flow demand with the corresponding maximum value in the input graph to make the model generalize to different input traffic demands.

\textbf{Model design}: Inspired by graph attention network \cite{brody2021attentive} and transformer \cite{vaswani2017attention}, the state-of-the-art deep learning mechanisms, we propose a graph-based attention mechanism to estimate the impact of target failure scenarios.
The overall architecture is shown in Fig. \ref{Fig:structure-GAT}.
The estimator consists of an input embedding layer, 5 local-global attention layers, and a readout layer.
For each local-global layer, we apply a multi-head attention mechanism to do the feature extraction for the graph-based input in both local and global views.
We apply the graph attention mechanism $(\mathbf{\Theta}, \mathbf{a})$ for the local view feature extraction as follow
\begin{equation}
    \mathbf{h}_i^*=\alpha_{i, i}\mathbf{\Theta} \mathbf{h}_j + \sum_{j\in \mathcal{N}(i)} \alpha_{i,j} \mathbf{\Theta} \mathbf{h}_j,
    \label{eq:attention-1}
\end{equation}
where $\mathcal{N}(i)$ denote the neighbor node set of $i$, $\mathbf{h}_i$ denote the state vector for node $i$, and the attention coefficients $\alpha_{i, j}$are computed as
\begin{equation}
    \alpha_{i, j} = \frac{\exp \left(\mathbf{a}^ \mathrm{T} {\rm LeakyReLU}(\mathbf{\Theta}[\mathbf{h}_i||\mathbf{h}_j])\right)}{\sum_{k\in \mathcal{N}(i)\cup\{i\}} \exp \left(\mathbf{a}^ \mathrm{T} {\rm LeakyReLU}(\mathbf{\Theta}[\mathbf{h}_i||\mathbf{h}_k])\right)}
    \label{eq:attention-2}
\end{equation}
With Eq. \eqref{eq:attention-1} and \eqref{eq:attention-2}, we let each node attend to its neighbors.
Such a one-hop attention process embeds the local graph structure information and benefits the link node to estimate the failure impact using the local view.
Besides local attention, we also have global attention in our model.
For the global view feature extraction, we do the attention among the link nodes with no neighbor constraints.  
Thus the link node could benefit from the global understanding when estimating failure impact.
The output of the global and local attention layer will be mixed with a fully connected layer.
Note that we apply residual and normalization mechanism in each global-local attention layer to support a stack of more layers.
Finally, the attention layers' aggregated representations of the failure combination nodes serve as the input for the readout layer to predict the failure impact
\begin{equation}
    y_x ={\rm Readout}(\mathbf{h}_x), x\in X.
    \label{eq:readout}
\end{equation}

\textbf{Generalization}: 
We argue that our proposed GAT-based failure impact prediction model intrinsically has good generalization since it can generalize to different network topologies and target sets of link failure scenarios without increasing or decreasing the number of model parameters. 
In addition, the input link state design, which embeds and normalizes the critical features of traffic demand and route,  enables the trained model to generalize to different patterns of traffic demands.
Further, we represent the original routing decision of a flow in the form of traversed paths and flow assignments, which can also be used to represent many other routing schemes (e.g., OSPF and ECMP) in real network scenarios.
Since the GAT-based model design is guided by the common insights of failure impact, we note that FERN has the potential to evaluate the failure impact of many real-world routing schemes.

\textbf{Computational efficiency}:
Algorithm based on deep learning usually requires much less computational overhead compared to the optimization model on large scale problems. 
The GAT-based model in FERN could generate the impact prediction of all the failure scenarios for a one-shot inference with little time overhead and memory use increase, enabling good computational efficiency and high scalability. 
We further evaluate the computational efficiency of FERN in \cref{sec:evaluation}. 

\subsection{Loss function}
\label{sec:loss}

In this section, we show the design of loss functions to train the GAT-based failure impact prediction model for a classification problem and a regression problem. 

\textbf{Critical Failure Classification}: 
We first show a relatively simple task to filter the possible critical failure scenarios. 
We divide the failure scenarios for a pair of given network topology, input traffic demand, and routing decision into three types:
\begin{itemize}
    \item \emph{Worst}: $\left\{x\in X | \frac{F_\theta(x, G, D, r)}{ \max_{x\in X} F_\theta(x, G, D, r)} \geq 0.95 \right\}$. 
    \item \emph{Significant}: $\left\{x\in X | 0.8 \leq \frac{F_\theta(x, G, D, r)}{ \max_{x\in X} F_\theta(x, G, D, r)} < 0.95 \right\}$
    \item \emph{Normal}: $\left\{x\in X | \frac{F_\theta(x, G, D, r)}{ \max_{x\in X} F_\theta(x, G, D, r)} < 0.8 \right\}$.
\end{itemize}
{\rev
The boundaries (i.e., 0.8 and 0.95) used to distinguish the three types of fault scenarios are determined based on the distribution of failure impacts in real world network topologies and the use cases in this work.
}
In this paper, we set the union of worst failure scenarios and significant failure scenarios as the set of critical failure scenarios $X_C$.
In such a task, the GAT-based model should output whether a failure scenario is critical.
Binary cross entropy (BCE) loss is a normal loss function for binary classification problems.
However, we find that with the BCE loss function, the trained model tends to give a conservative prediction.
That is, the trained model likely misses some critical failure scenarios to fit the major normal failure scenarios.
Such a conservative model may suffer a terrible performance degradation under some application scenarios. 
In this work, we define weighted BCE loss, which calculates the loss for each type of failure scenario independently and eliminates the quantitative advantage of normal failure scenarios in original BCE loss. 
The weighted MSE loss function is defined as follow
\begin{equation}
    \mathcal{L}_{WBCE} = -\sum_{t=1}^3 \frac{1}{n_t}\sum_{j=1}^{n_t} \left[y_j^t\log \hat{y}_j^t  + (1 - y_j^t)\log (1 - \hat{y}_j^t)\right]
\end{equation}
where $n_t, y^t_j, \hat{y}^t_j$ denote the number of the failure scenarios of type $t$, prediction value of $j$-th failure scenario of type $t$, and ground truth value of $j$-th failure scenario of type $t$, respectively.

\textbf{Failure Impact Regression}:
To further model the failure impact, we also introduce a failure impact regression task in this paper.
In such a task, the GAT-based model learns to output the MLU increase of the target failure scenarios as shown in Eq. \eqref{eq:impact-def-FERN}. 
Similar to the classification task, we define the weighted mean square error loss function as
\begin{equation}
    \mathcal{L}_{WMSE} = \sum_{t=1}^3 \frac{1}{n_t}\sum_{j=1}^{n_t}(y_j^t-\hat{y}_j^t)^2
\end{equation}
In contrast to the critical failure classification problem, learning to accurately model the impact of all the target failure scenarios is much more challenging.
In this paper, we focus on critical failure detection, so moderate errors to the normal failure scenarios are acceptable.
Thus we propose a self-adaptive failure scenario sampling scheme to dismiss some well-trained failure scenarios during training.
In particular, the worst failure scenarios are kept for each training epoch, the significant failure scenarios are likely to be kept with a probability dependent on the relative error in the last training epoch, and the normal failure scenarios are likely to be dismissed once the relative error is acceptable in last training epoch.
With such a failure scenario sampling scheme, the model could focus on the untrained, difficult failure scenarios and converge to a good strategy faster.

In order to further improve the model capability in detecting critical failure scenarios, we propose an additional loss function. This additional loss function encourages the model to distinguish the critical failure scenarios from normal ones even under situations where accurately predicting failure impact is difficult. The additional loss function  is defined as follows 

\begin{equation}
\begin{aligned}
    &\mathcal{L}_{add} = \\
    &{\rm RELU}\left(\sum_{j=1}^{n_1}(y_j^1-\hat{y}_j^1)^2  - \frac{1}{n_2 + n_3} \sum_{t=2}^3 \sum_{j=1}^{n_t}(y^{t}_j-\hat{y}_j^{t})^2 \right).
\end{aligned}
\end{equation}
Such an additional loss function encourages the model to output higher prediction failure impact for the critical failure scenarios than the normal ones. 
Note that the additional loss function above can be easily implemented and optimized with a gradient decent algorithm. 
We simply add up the two loss function as the final loss function for the failure impact regression problem.

\subsection{Training and Inference}
In the training phase, we have to balance the trade-off between generalization and accuracy. 
Training a model for one or a small set of topologies will obtain good performance in prediction accuracy but at the cost of poor generalization to unseen topologies. 
However, in a real-world application scenario, we may not always have enough time to obtain enough training data for specific network topology, especially for large wide area network (WAN) and highly dynamical overlay network.
On the other hand, training a model which generalizes to any application scenario requires a large amount of data, large-scale model parameters, and high computational overhead. 
Even so, it is still difficult to guarantee that the model can achieve high accuracy in all scenarios.

In this section, we propose a two-stage learning framework to balance both the generalization and accuracy of this work.
Generally, we train a general model over a large dataset in the first training phase, and we could further improve the model accuracy with a small task-specific dataset in the second training phase.
With the trained GAT-based FERN model, we could evaluate the failure impact and choose the critical failure scenarios efficiently.

\textbf{First training phase}: 
We first train a general model with a dataset comprised of numerous randomly generated and real-world network topologies, randomly generated link capacities, and randomly generated traffic matrices. 
We find that a larger dataset with more network topologies and link capacities makes the model obtained by the first training phase generalize better. 
At the same time, a larger training dataset also means that the model takes longer to train and is more difficult to converge.
After the first training phase, we obtain a general model that can achieve good performance for most network topologies and traffic demands. 

\textbf{Second training phase}: 
If the general model trained in the first phase does not meet the accuracy requirements of the application scenario,
we could apply a second training phase for the task specifically. 
In this phase, we train the model with a small task-specific dataset for a few extra steps to obtain a reliable and accurate model.
The second training phase could be applied to unseen large topologies, real-world traffic matrices, and non-optimal routing policies.   
This training process improves the algorithm's generalization to those challenging application scenarios, which are hard to be covered in the first training phase.

\textbf{Critical Failure Inference}: 
With the trained critical failure classification and failure impact regression model, we could choose the critical failure scenarios efficiently. 
Generally, we apply the results of the classification model to filter possible critical failure scenarios.
We rank the possible critical failure scenarios with the output failure impact prediction by regression model and build up the critical failure set.

\section{FERN for Robust Network Design}
\label{sec:fault-tolerant-network-design}

In this section, we show how to use FERN to formulate and solve three important robust network design problems, namely network robust validation \cite{chang2017robust}, network upgrade optimization  \cite{chang2017robust,ahuja2021capacity} and fault-tolerant traffic engineering \cite{wang2010r3}.
We recast the three robust network design problems using the proposed failure impact evaluation function $F_\theta$ and show how to apply the FERN-predicted critical failure scenarios to efficiently solve the recast problems.
We also give some performance guarantee to the FERN-based robust network design algorithm in Theorem \ref{theory:network-upgrade}.

In particular, \emph{network robust validation} needs to find the worst failure scenarios that have the most severe impact on certain network performance objectives. We can directly use the proposed function approximation $F_\theta(x,G,D,r)$ to identify such critical failures for scalable network validation. 
Next, we consider \emph{network upgrade optimization}, which aims to minimize the cost of necessary link capacity upgrades subject to network congestion constraints under link failures. 
We show that by pruning the non-essential failure scenarios considered for network upgrades, we can substantially reduce the problem space of such network upgrade optimization. 
Finally, we also design a \emph{fault-tolerant traffic engineering} algorithm using FERN. 
Instead of enumerating all failure scenarios, it features a nearly-optimal rerouting strategy over a small set of critical failure scenarios while optimizing basic load balancing objectives.

We hasten to emphasize that all three use cases are based on FERN's function approximation $F_\theta(x,G,D,r)$ for scalable failure evaluation. 
It allows robust network design problems to be formulated with respect to only a small subset of failure scenarios that have significant impact on robustness. 
By providing a GAT-based function approximation, FERN supports a broad range of robust network optimization problems.

\subsection{Network robust validation}
\label{sec:robust-validation}

Robust validation \cite{chang2017robust} establishes the worst case performance of a network over all possible failure scenarios. Let $r(d,e)$ be the traffic of demand $d$ assigned to link $e$ and $T(d, v)$ be the known traffic demand $d$ originating from node $v$. A classic robust network validation problem under optimal MCF routing scheme can be formulated as a two stage max-min optimization finding the worst case failure $x$ to maximize congestion $U$ over optimal routing strategies $r$:
\begin{equation}
    \begin{gathered}
\max_{x\in X} \min_{r} ~U 
    \\
     s.t. \\ 
     \sum_{d} r(d, e) \leq  U \cdot C_{e}(1-x_{e}), \forall e\in E  \\
    \sum_{e_{src}=v}r(d, e) - \sum_{e_{dst}=v}r(d, e) = T(d, v), 
     \forall v \in V, d\in D\\
    r(d, e) \geq 0, ~ \forall d \in D,e \in E 
    \end{gathered}
    \label{eq:robust-validation}
\end{equation}
where $x_e=1$ (thus resulting in zero capacity) if link $e$ fails and $x_e=0$ otherwise. The optimization is subject to standard congestion and traffic constraints.


The max-min optimization in Eq. \eqref{eq:robust-validation} has been shown to be NP-complete \cite{chang2017robust}. Due to the combinatorial nature of multiple link failures, existing works like \cite{chang2017robust} often solve the problem via approximation approaches to balance optimality and tractability. 
However, it may cause performance degradation and becomes infeasible in more complex networks such as today's hundred-node large WAN topology.

Making use of FERN, this network robust validation problem can be readily reformulated through $F_\theta$, i.e.,
\begin{equation}
    \begin{gathered}
        \max_{x\in X} F_\theta(x, G, D, r^o)
    \end{gathered}
    \label{eq:robust-validation-F}
\end{equation}
where $r^o$ is the optimal route and reroute strategy embedded in our GAT-based model. 
The worst case failures solving Eq. \eqref{eq:robust-validation-F} are exactly the most critical failure scenarios with the highest impact (i.e., PR) in $F_\theta(x, G, D, r^o)$. This neural network function approximation allows us to quickly sift through different failure scenarios and focus robust validation on a very small set of critical failure scenarios identified by FERN (as opposed to all possible failures). 
We also note that the perspective to speed up robust network validation by locating critical failure scenarios with FERN is also applicable to other routing frameworks (e.g., OSPF), which we show further in \cref{sec:evaluation}.

\subsection{Network upgrade optimization}
\label{sec:network-update}

Network upgrade problems minimize the total cost of necessary link bandwidth increases in order to satisfy network availability constraints under future traffic demands and potential failures. It is an integral part of WAN network planning problems considered in \cite{chang2017robust, ahuja2021capacity, zhu2021network}.
Let $r_x(d, e)$ be the traffic of demand $d$ assigned to link $e$ under failure scenario $x\in X$, and $a_e$ be the added bandwidth on link $e$. We consider a typical network upgrade problem with a linear cost function, formulated as an integer linear programming (ILP):
\begin{equation}
    \begin{gathered}
    \min_{r, a}~ \sum_{e\in E} a_e            \\
        s.t., \\
          \sum_{d} r_x(d, e)  \leq U\cdot (C_e + a_e)(1 - x_e), \forall e \in E, x \in X  \\
          \sum_{e_{src}=v}r_x(d, e) - \sum_{e_{dst}=v}r_x(d, e) = T(d, v),
         ~\forall v \in V,  d\in D, x \in X \\
          r_x(d, e) \geq 0, \forall d\in D, e \in E, x\in X, \\
         U\le 1, \ a_e \geq 0, \forall e\in E
    \end{gathered}
    \label{eq:network-upgrade}
\end{equation}
where $U\le 1$ ensures that the network is congestion free. 
The problem aims to find the least expensive bandwidth upgrades to ensure that the network is congestion free under given traffic demands $D$ and link failure scenarios $X$. We note that the numbers of decision variables and constraints both grow quickly with the number of failure scenarios in $X$. Thus, finding the optimal upgrade strategy is difficult as the problem space becomes increasingly complex when multiple link failures are considered.

Now we leverage FERN to recast the network upgrade problem. Let $\Delta G$ denote the set of bandwidth upgrades on network $G$. Using our definition of $F_{\theta}$ in Eq. \eqref{eq:impact-def-FERN}, the congestion-free constraint in Eq. \eqref{eq:network-upgrade} with respect to different failures $x\in X$, i.e., $MLU(x, G+\Delta G, D, r^o)\le 1$, can be rewritten as $\forall x \in X$:
\begin{equation}
    \begin{gathered}
    F_{\theta}(x, G+\Delta G, D, r^o) \cdot MLU(\phi, G+\Delta G, D, r^o) \leq 1,
    \end{gathered}
    \label{eq:network-upgrade-F}
\end{equation}
where $MLU(\phi, G+\Delta G, D, r^o)$ is the congestion without failures and can be easily computed. 

It shows that for minimizing an upgrade cost $cost(\Delta G)$, only the worst failure scenario with maximum failure impact  $F_{\theta}(x, G+\Delta G, D, r^o)$ needs to be validated in Eq. \eqref{eq:network-upgrade-F}. However,  $F_{\theta}(x, G+\Delta G, D, r^o)$ depends on the upgrade policies $\Delta G$, which are hard to enumerate. To this end, we show that critical failure scenarios in the original network $G$ can provide insights and help circumvent this challenge. Let $x^w$ be the worst failure scenario for topology $G$ and traffic demand $D$ under optimal route $r^o$. The result is formally stated in the following theorem.
\begin{theorem}
     Eq. \eqref{eq:network-upgrade-F} holds for all failures $x\in X$, if it holds on the following set of critical failures ${X}_C$:
     ${X}_C = \left\{x \ \Big| \ F_{\theta}(x, G, D, r^o) \geq \frac{F_\theta(x^w, G, D, r^o)}{MLU(x^w, G, D, r^o)} \right\}$.
    \label{theory:network-upgrade}
\end{theorem}
The proof of Theorem \ref{theory:network-upgrade} is shown in \cref{sec:proof-network-upgrade}. 
The theorem indicates that for network upgrade optimization, we only need to consider critical failures with impact larger than $\frac{F_\theta(x^w, G, D, r_o)}{MLU(x^w, G, D, r_o)}$, which are computable on topology $G$ rather than $G+\Delta G$.   
In \cref{sec:evaluation}, we show it enables significant pruning of the problem space. 

\subsection{Fault-tolerant traffic engineering}
\label{sec:fault-tolerance-TE}
We consider a link-based resilient routing problem similar to \cite{wang2010r3} to demonstrate how to employ FERN for fault-tolerant traffic engineering.
In particular, when a link $l$ fails (i.e., $x_l=1$), we find decisions $ \hat{r}(l, e)$ to reroute the traffic on $l$ onto links $e\in E$, such that the network remains congestion free under all failure scenarios $x\in X$.
This fault-tolerant TE problem can be formulated as:
\begin{equation}
\begin{gathered}
    \min_{r,\hat{r}}~ U \\
    s.t. \\ 
     \sum_{e_{src}=v}r(d, e) - \sum_{e_{dst}=v}r(d, e) = T(d, v), \forall v \in V, d\in D \\
     r(d, e) \geq 0, \forall d \in D, e \in E  \\
     \sum_{e_{src} =v}\hat{r}(l, e) - \sum_{e_{dst}=v}\hat{r}(l, e) = T(l, v), \forall v \in V, l \in E  \\
     \hat{r}(l, e) \geq 0,  \forall l\in E, e \in E  \\
      \sum_{d} r(d, e) + \sum_{l} \hat{r}(l, e)  x_l  \leq UC_e, \forall e \in E, x \in X  \\
\end{gathered}
      \label{eq:R3-origin}
\end{equation}
where $T(l, v)$ is the extra traffic demand for failed link $l=(i,j)$, from node $i$ to $j$ with traffic volume equal to link capacity $C_l$. The optimal reroute decisions $\hat{r}(l, e)$ in Eq. \eqref{eq:R3-origin} allows an online reconfiguration for each failed link while ensuring that the network remains congestion free under any potential failures in $X$.
    
Again, the key challenge here is the huge problem space that stems from the combinatorial nature of multiple failures. 
Existing solutions like R3 \cite{wang2010r3} limit the failure set $X$ to $f$ simultaneous link failures and then relax the problem for the tractability.
While the relaxed problem can be computed through linear programming, it is known that R3 cannot achieve optimal rerouting and fails to cope with certain unresolved failure scenarios \cite{chang2019lancet}.

It is easy to see that the congestion objective in fault-tolerant TE (i.e., $ MLU(x, G, D, r)$) is the product of failure impact $F_{\theta}(x, G, D, r)$ and failure-free congestion $MLU(\phi, G, D, r)$ following the definition in Eq.\eqref{eq:impact-def-FERN}. We obtain an equivalent formulation using FERN:
\begin{equation}
        \min_{r\in \mathcal{R}} \max_{x\in X} F_{\theta}(x, G, D, r) \cdot MLU(\phi, G, D, r),
    \label{eq:Robust-TE-F}
\end{equation}
where 
the rerouting decisions $r$ are limited to a set of permissible strategies $\mathcal{R}$ and is optimized over all failure cases $x\in X$. It finds the best permissible rerouting decisions that optimize the worst case performance under possible failures.


Eq. \eqref{eq:Robust-TE-F} indicates that we should select a reroute from permissible strategies $\mathcal{R}$ which achieves performance as close as possible to theoretical optimal rerouting $r^o$, especially under the critical failure scenarios.   
Using FERN, we propose a critical failure scenario oriented robust traffic engineering scheme to solve the problem.
We optimize worst case congestion under the critical failure scenarios selected by FERN while ensuring a good performance under normal failure scenarios with a simple load balancing objective.
Compared to the original problem shown in Eq. \eqref{eq:R3-origin}, we modify the objective function to combine the critical failure performance and normal load balance objective as
\begin{equation}
    \min_{r,\hat{r}}~ U_C + \frac{1}{|E|}\sum_{\mathcal{E}} U'_\mathcal{E},
\end{equation}
where $U_C$ is worst case congestion under critical failure scenarios in $X_C$ and $\frac{1}{|E|}\sum_{\mathcal{E}} U'_\mathcal{E}$ is expected congestion under single link failure scenarios in $X_1$. 
The full optimization problem formulation is shown in \cref{sec:Robust-TE-appendix}.
We note that the new optimization problem only considers $|X_1|+|X_C|$ failure scenarios, which are much smaller than the original problem.

The newly proposed optimization problem significantly reduces the number of constraints of the original fault-tolerant TE problem, but there is no guarantee that the generated routing decision is congestion free in all the failure scenarios.
However, although solving the problem in Eq. \eqref{eq:R3-origin} is time consuming, it is much easier to validate the performance of a given solution $r,\hat{r}$.
Thus we can iteratively validate, add new bad failure scenarios to the critical failure set $X_C$, and solve the optimization problem again until the solution is congestion free for all scenarios.
In \cref{sec:evaluation} we show that our proposed fault tolerant TE algorithm using FERN could achieve similar performance to the original optimization problem in Eq. \eqref{eq:R3-origin} while reducing computational overhead significantly.


\section{evaluation}
\label{sec:evaluation}

In evaluation, we aim to answer the following key questions:
\begin{itemize}
    \item How is the capability of FERN in failure evaluation and critical failure detection (\cref{sec:model-performance})?
    \item How is the generalizability of FERN to unseen traffic matrices and network topologies (\cref{sec:model-performance})?
    \item How is the performance of using FERN in robust network design (\cref{sec:use-case})?
\end{itemize}

\subsection{Experiment setup}

\begin{table}[ht]\footnotesize
        \caption{Dataset information for optimal MCF routing.}
        \label{tab:dataset-info}
        \centering
        \scalebox{0.8}{
        \begin{tabular}{cccc}
            \toprule
            Dataset & Topology number & Topology scale & TM number\\
            \midrule
            BRITE-small-train & 90 & 6-15 & 1000\\
            Topology-zoo-small-train & 69 & $<$ 20 & 1000 \\
            Topology-zoo-middle-train & 78 & 20 - 80 &10 \\
            \midrule
            BRITE-small-test1 & 90 & 6-15 & 100 \\
            Topology-zoo-small-test1 & 69 & $<$ 20 & 100 \\
            Topology-zoo-middle-test1 & 78 & 20 - 80 &1 \\
            \midrule
            BRITE-small-test2 & 10 & 6-15 & 100 \\
            Topology-zoo-small-test2 & 10 & $<$ 20 & 100 \\
            Topology-zoo-middle-test2 & 20 & 20 - 80 &1 \\
            \midrule
            Topology-zoo-large-test & 6 & 80 - 114 & 1 \\
            Abi-realTM-test & 1 & 11 & 4032 \\
            GEA-realTM-test & 1 & 23 & 1344 \\
            \bottomrule
            \end{tabular}}
    \end{table}

We implement our GAT-based failure impact prediction model with PyTorch Geometric \cite{Fey/Lenssen/2019} and run the training and evaluation process on a server with two Intel Xeon Gold 6230R CPUs and two RTX 2080S GPUs. 
We consider single and double simultaneous link failures in evaluation.
The experimental settings for optimal MCF routing is shown below. Similar experimental settings are used for OSPF routing, refer to \cref{sec:FERN-OSPF} for details.

\textbf{Dataset}:
The overall information of the datasets is shown in Table \ref{tab:dataset-info}. 
We train and evaluate our model with both 100 randomly generated small topologies using BRITE \cite{medina2001brite} and real-world topologies from topology zoo \cite{topology-zoo}.
We eliminate the one-degree nodes in the topologies of topology zoo.
In order to test the performance of FERN on unseen topologies, we split the topologies above into two parts.
The major part are placed in \emph{train} and \emph{test1} dataset while the others are in \emph{test2} dataset.  
We randomly generate several traffic demand matrices (TM) for each topology using the gravity model \cite{geng2021distributed}. 
We combine a topology, a traffic matrix, and a set of link capacities as a piece of training data.
In order to simulate the heterogeneous link capacities, the link capacity of each link is randomly selected from 1/4, 1/2, 3/4, and 1 for each data piece.
We implement the plain MCF optimization model with gurobi to calculate the failure impact under each single and double simultaneous link failure of each data piece.
For middle and large topologies, we only generate a small number of data pieces for each topology because of the huge computational overhead.
We also use the real-world traffic matrices of Abi and GEANT topology measured by \cite{Abilene2004,uhlig2006providing} in our evaluation.

\begin{figure*}[t]\footnotesize
\centering
\begin{minipage}[t]{0.3\textwidth}
\centering
\includegraphics[width=\linewidth]{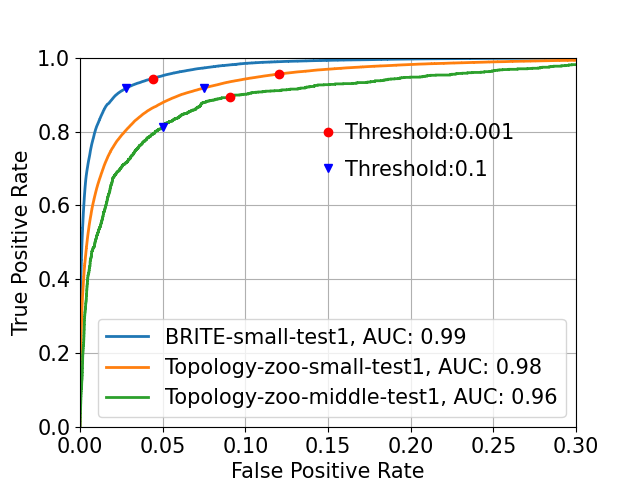}
\captionsetup{font={footnotesize}}
\caption{\footnotesize Accuracy (ROC curves) of FERN classification model on seen topologies.}
\label{Fig:P1-ROC-test1}
\end{minipage}
\hspace{5mm}
\begin{minipage}[t]{0.3\textwidth}
\centering
\includegraphics[width=\linewidth]{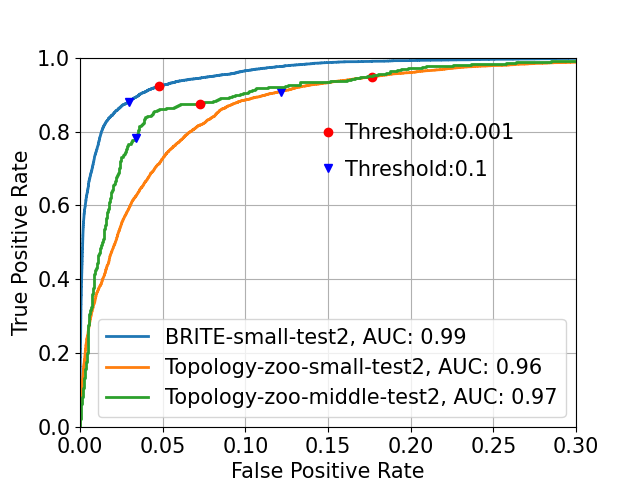}
\captionsetup{font={footnotesize}}
\caption{\footnotesize Accuracy (ROC curves) of FERN classification model on unseen topologies.}
\label{Fig:P1-ROC-test2}
\end{minipage}
\hspace{5mm}
\begin{minipage}[t]{0.3\textwidth}
\centering
 \includegraphics[width=\linewidth]{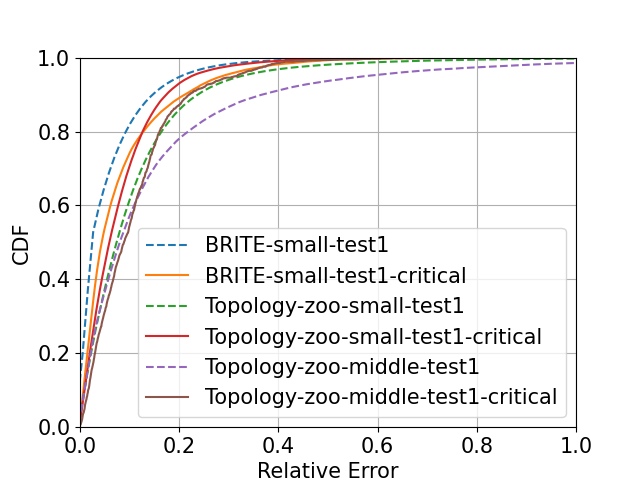}
\captionsetup{font={footnotesize}}
\caption{\footnotesize Relative error of FERN regression model on seen topologies.}
\label{Fig:P1-mre-test1}
\end{minipage}
\hspace{5mm}
\begin{minipage}[t]{0.3\textwidth}
\centering
 \includegraphics[width=\linewidth]{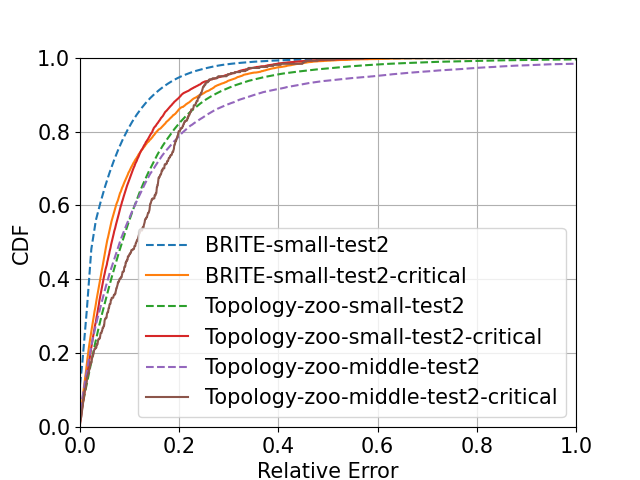}
\captionsetup{font={footnotesize}}
\caption{\footnotesize Relative error of FERN regression model on unseen topologies.}
\label{Fig:P1-mre-test2}
\end{minipage}
\hspace{5mm}
\begin{minipage}[t]{0.3\textwidth}
\centering
\includegraphics[width=\linewidth]{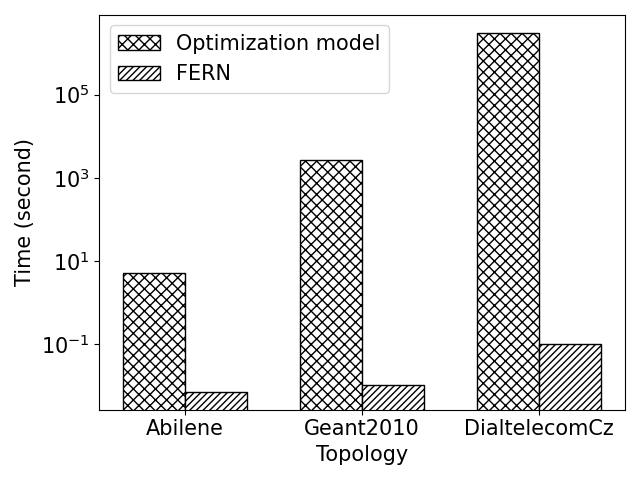}
\captionsetup{font={footnotesize}}
\caption{\footnotesize Failure evaluation time efficiency of FERN.}
\label{Fig:Computing-time}
\end{minipage}
\hspace{5mm}
\begin{minipage}[t]{0.3\textwidth}
\centering
\includegraphics[width=\linewidth]{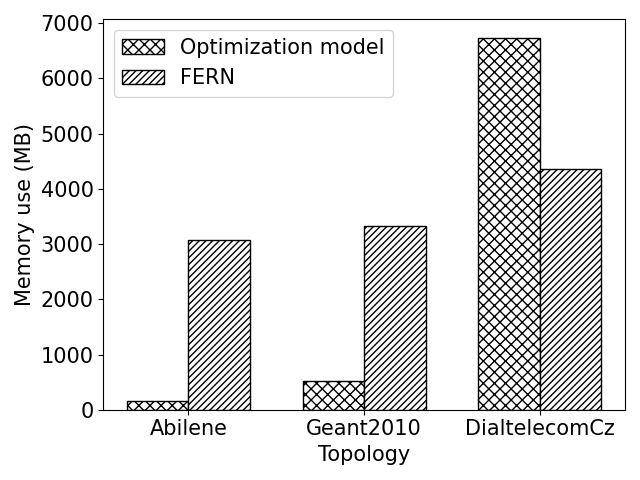}
\captionsetup{font={footnotesize}}
\caption{\footnotesize Failure evaluation memory use of FERN.}
\label{Fig:Computing-memory}
\end{minipage}
\end{figure*}

\textbf{Training}:
We train a general classification model and a general failure impact regression model over the joint dataset consisting of the three training datasets shown in Table \ref{tab:dataset-info}.
The details of the model hyper-parameter settings are introduced in \cref{sec:hyper-parameter}, and the details of the training process are introduced in \cref{sec:training}.
We note that the first training phase takes about ten days on the server.
In addition to optimal MCF routing, we also train a general model for the OSPF routing based on the dataset generated with the OSPF routing scheme. 

\textbf{Robust validation}: 
With the datasets above, we can easily obtain the ground truth failure impact for robust validation. 
We compare the prediction accuracy of the worst-case network performance and computational efficiency between FERN-based algorithm and the plain optimization approach, i.e., solving an MCF problem for each enumerated failure scenario. 

\textbf{Network upgrade}:
We solve the network upgrade optimization problem in Eq.\eqref{eq:network-upgrade} with gurobi optimizer as an optimal baseline,{\rev which also refers to \cite{chang2017robust}}.
We compare the augmented network performance and computational time efficiency between FERN-based algorithm and the plain optimization approach.

\textbf{Fault-tolerant TE}:
We solve the link-based resilient TE problem in Eq.\eqref{eq:R3-origin} with gurobi optimizer as an optimal baseline, {\rev which also refers to \cite{wang2010r3}}.
We compare the performance of generated routing and rerouting decision under online reconfiguration in \cite{wang2010r3} as well as computational time efficiency between FERN-based algorithm and plain optimization approach.
We note that the approximated approach in R3 \cite{wang2010r3} is not suitable for the problem since protecting all the 2 simultaneous link failures is not feasible in many topologies.

\subsection{Failure evaluation}
\label{sec:model-performance}

In this section, we evaluate the accuracy, generalizability, and computational efficiency of our proposed GAT-based failure impact prediction algorithm in FERN.

\begin{table*}[ht]\footnotesize
    \caption{Performance, i.e., mean relative error (MRE), MRE for critical failure scenarios (MRE.C) and area under ROC curve (AUC),  of FERN over large topologies in topology zoo.}
    \label{tab:performance-large-topologies}
    \centering
    \scalebox{0.8}{
    \begin{tabular}{c|ccc|ccc|ccc|ccc|ccc|ccc}
        \toprule
        \multirow{2}*{\textbf{Model}} & \multicolumn{3}{|c}{\textbf{Colt}} & \multicolumn{3}{|c}{\textbf{VtlWavenet2008}} & \multicolumn{3}{|c}{\textbf{VtlWavenet2011}} & \multicolumn{3}{|c}{\textbf{DialtelecomCz}} & \multicolumn{3}{|c}{\textbf{Interoute}}& \multicolumn{3}{|c}{\textbf{Ion}} \\
        \cline{2-19}
        & \footnotesize \textbf{MRE} & \footnotesize  \textbf{MRE.C} & \footnotesize \textbf{AUC} & \footnotesize \textbf{MRE} & \footnotesize \textbf{MRE.C} & \footnotesize \textbf{AUC}  & \footnotesize \textbf{MRE} & \footnotesize \textbf{MRE.C} & \footnotesize \textbf{AUC} & \footnotesize \textbf{MRE} & \footnotesize \textbf{MRE.C} & \footnotesize \textbf{AUC} & \footnotesize \textbf{MRE} & \footnotesize \textbf{MRE.C} & \footnotesize \textbf{AUC} & \footnotesize \textbf{MRE} & \footnotesize \textbf{MRE.C} & \footnotesize \textbf{AUC} \\
        \midrule
        P1 & 0.20 & 0.15 & 0.99 & 0.19 & 0.12 & 0.72 & 0.24 & 0.08 & 0.76 & 0.35 & 0.09 & 0.82 & 0.31 & 0.37 & 0.83 & 0.22 & 0.23 & 0.96 \\
        P1 + P2 & \textbf{0.13} & 0.17 & 0.97 & \textbf{0.12} & \textbf{0.08} & \textbf{0.95} & \textbf{0.12} & \textbf{0.05} & \textbf{0.95} & \textbf{0.12} & \textbf{0.06} & \textbf{0.98} & \textbf{0.08} & 0.41 & \textbf{0.90} & \textbf{0.11} & \textbf{0.08} &  \textbf{1.00} \\
        \bottomrule
        \end{tabular}}
\end{table*}

\begin{table}[ht]\footnotesize
    \caption{Performance of FERN, i.e., MRE, MRE.C and AUC, over real-world traffic matrices.}
    \label{tab:performance-real-world-TM}
    \centering
    \scalebox{0.8}{
    \begin{tabular}{c|ccc|ccc}
        \toprule
        \multirow{2}*{\textbf{Model}} &\multicolumn{3}{|c}{\textbf{Abi}} & \multicolumn{3}{|c}{\textbf{GEANT}}  \\
        \cline{2-7}
        & \footnotesize \textbf{MRE} & \footnotesize  \textbf{MRE.C} & \footnotesize \textbf{AUC} & \footnotesize \textbf{MRE} & \footnotesize \textbf{MRE.C} & \footnotesize \textbf{AUC} \\
        \midrule
        P1 & 0.32 & 0.43 & 0.63 & 0.16 & 0.55 & 0.80 \\
        P1+P2 & \textbf{0.03} & \textbf{0.03} & \textbf{0.95} & \textbf{0.08} & \textbf{0.10} & \textbf{0.93} \\
        \bottomrule
        \end{tabular}}
\end{table}

\textbf{FERN classification model}:
We show the accuracy of the FERN classification model after the first training phase (P1) in Fig. \ref{Fig:P1-ROC-test1} and Fig. \ref{Fig:P1-ROC-test2}.
We can find that on all the six datasets, the FERN classification model achieves more than 0.87 (0.78) true positive rate, i.e., the ratio of critical failure scenarios detected by FERN, with a threshold 0.001 (0.1).
FERN achieves the best prediction accuracy on the BRITE-small dataset with the largest area under curve (AUC) value. 
Moreover, FERN achieves a higher true positive rate on small-scale topologies than on middle-scale topologies in topology zoo at the cost of selecting more non-critical failure scenarios.
In general, the high true positive rate combined with low false positive ratios indicate that FERN could learn a general and accurate policy to classify the critical failure scenarios for different topology scales.
Moreover, similar performances on seen and unseen topologies in FERN's training phase indicate that FERN has good generalizability towards unseen network topologies and has great potential to be applied in fast-evolving network environments.

\textbf{FERN regression model}:
We evaluate the FERN failure impact regression model after P1 training on the relative error between the predicted failure impact and the ground-truth value.
The results are shown in Fig. \ref{Fig:P1-mre-test1} and \ref{Fig:P1-mre-test2}.
Since we focus on the prediction accuracy of the critical failure scenarios, we also show the relative error toward the critical failure scenarios.
We can find that more than 80\% of the failure scenarios have relative errors less than 0.2, and the relative error of the critical failure scenarios is better bounded by 0.6.
We also note that the relative error of critical failure scenarios is better than normal failure scenarios when the problem becomes complicated, i.e., for larger-scale real-world topologies.
Such results show the effect of our loss function design as mentioned in \cref{sec:loss}.
Thus we show that the FERN regression model could predict the impact of most failure scenarios accurately and are ready to predict critical failure scenarios.


\textbf{Generalization to large topologies}:
We further evaluate FERN over large topologies.
We evaluate the performance of FERN after the first training phase (P1) and after an extra second training phase (P2).
For the extra P2 training, we randomly select 10\% \emph{Normal} and 20\% \emph{Critical} failure scenarios of a large topology for training and evaluate the model over all failure scenarios.
In P2 training, we apply an extra 50 training steps for the FERN classification model and regression model.
We show the results in Table \ref{tab:performance-large-topologies}.
We can find that after P1 training, FERN  could classify the critical failure scenarios and predict the impact of critical failure scenarios, although it never learns from such large-scaled topologies.
The results indicate that FERN has learned some general strategies for failure evaluation regardless of the topology scale. 
We note that P2 training brings significant performance gains to FERN over the six large topologies.
Further, for Interoute topology, the FERN regression model fails to model the failure impact for the worst failure scenarios accurately.
We point out that an important reason for this phenomenon is that there are only 3 critical failure scenarios for Interoute, and only one of them is used for P2 training. Collecting more critical failure scenarios in P2 training can significantly improve model performance.
Thus we demonstrate the amazing potential of FERN to generalize to much larger WAN network in the future.

\textbf{Generalization to real-world traffic matrices}:
We also evaluate FERN over unseen real-world traffic matrices measured in two weeks.
The results are shown in Table \ref{tab:performance-real-world-TM}.
We can find that FERN suffers a significant performance degradation on both Abi and GEANT with real-world traffic matrices.
However, we show that FERN could learn to generalize to such traffic matrices with a 100-step second training phase.
In particular, for each topology, we additionally train FERN over the traffic matrices measured in the first week and then evaluate the trained model over the traffic matrices measured in the second week.
We find that with P2 training, which only takes a few minutes, FERN obtains a significant performance improvement over the testing topology and traffic matrices.
We note that P2 training can be seen as a few-shot learning, which balances the generalization and domain-specific performance of FERN.

\textbf{Computational efficiency}:
To show the performance gain of our approach in computational overhead, we compare the memory use and time overhead of failure evaluation between FERN and the plain optimization approach, which solves an MCF problem for each candidate failure scenario.
The results are shown in Fig. \ref{Fig:Computing-time} and \ref{Fig:Computing-memory}.
We can find that the time overhead to enumerate all of the failure cases and solve them with an optimization model becomes extremely high when the network topology is large. 
On the contrary, FERN shows obvious gains in time efficiency, especially for large topologies.
Moreover, FERN shows less increase in memory use compared to the plain optimization approach, indicating good scalability. 

\textbf{FERN for OSPF routing scheme}: Similar to optimal MCF routing, we evaluate the accuracy and generalizability of the GAT-based failure impact evaluation function on the OSPF routing scheme. We highlight that both mean relative errors (MRE) and MRE for critical failure scenarios (MRE.C) for the regression model are less than 0.15 on all the test datasets shown in Table. \ref{tab:dataset-info-OSPF}, including a dataset consisting with 11 unseen large-scale topologies with up to 166 nodes.  More details of the evaluation results are shown in \cref{sec:FERN-OSPF}.   

\subsection{Robust network design}
\label{sec:use-case}

\begin{figure*}[htbp]\footnotesize
\centering
\begin{minipage}[t]{0.3\textwidth}
\centering
\includegraphics[width=\linewidth]{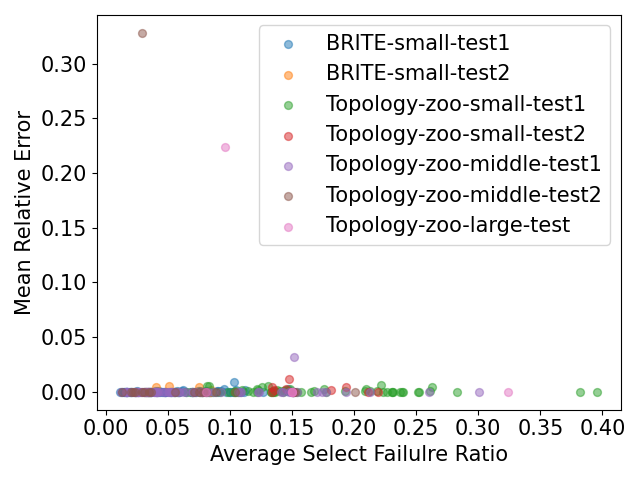}
\captionsetup{font={footnotesize}}
\caption{\footnotesize Performance of FERN-based robust validation.}
\label{Fig:robust-validation}
\end{minipage}
\hspace{5mm}
\begin{minipage}[t]{0.3\textwidth}
\centering
\includegraphics[width=\linewidth]{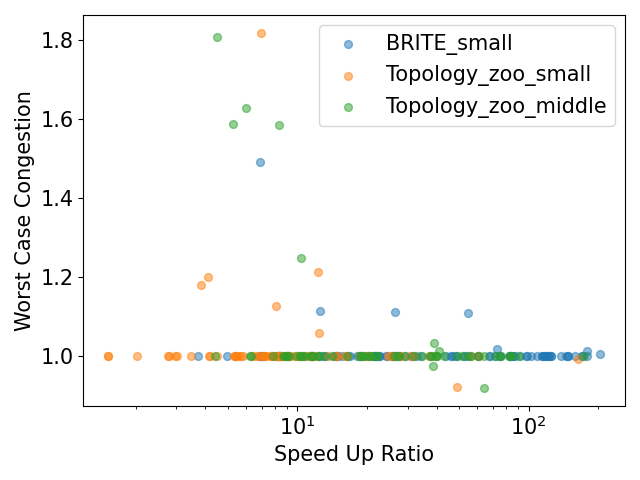}
\captionsetup{font={footnotesize}}
\caption{\footnotesize Performance of FERN-based network upgrade optimization.}
\label{Fig:network-upgrade}
\end{minipage}
\hspace{5mm}
\begin{minipage}[t]{0.3\textwidth}
\centering
\includegraphics[width=\linewidth]{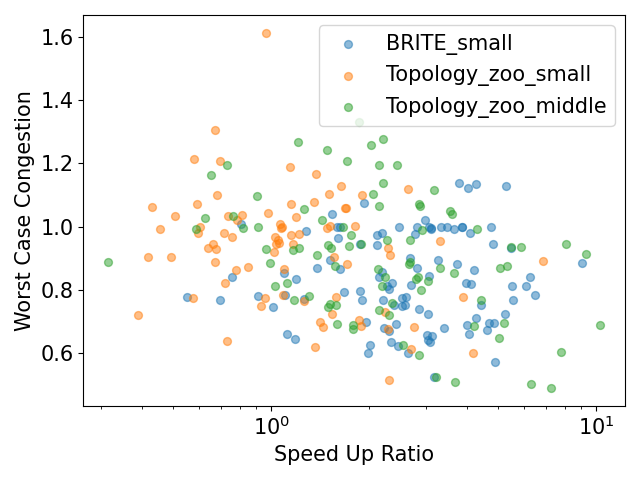}
\captionsetup{font={footnotesize}}
\caption{\footnotesize Performance of FERN-based fault tolerant TE.}
\label{Fig:robust-TE}
\end{minipage}
\vspace{-3mm}
\end{figure*}

In this section, we illustrate how FERN enhances the scalability of robust network design problems.
We generate the critical failure set $X_C$ using FERN model after P1 training.

\textbf{Robust network validation}:
Fig. \ref{Fig:Computing-time} shows the unacceptable time overhead for the plain optimization approach to solve the robust validation problem for large network topology.
We then evaluate the performance of applying FERN in robust validation problems.
The results are shown in Fig. \ref{Fig:robust-validation}, where each dot in the figure represents a topology in the testing dataset.
We note that when applying FERN to reduce the enumeration in robust validation problem, the time reduction is proportional to the number of failure scenarios selected.
We can find that for most topologies, the worst failure scenarios are included in the selected failure scenarios by FERN, and in almost all the test cases, the relative error is bounded by 0.05.
At the same time, FERN selects less than 15\% failure scenarios for most middle and large topologies, while for some small topologies, the ratio of selected failure scenarios could be slightly more than 15\%.
Moreover, we can find that the ratio of selected critical failure scenarios becomes smaller when the topology scale is larger, indicating good scalability of the FERN-based robust validation algorithm.  
We also note that only less than 1\% of failure scenarios are critical failure scenarios in many middle and large topologies.
So we can expect that a more accurate FERN model after P2 training will bring more time reduction.
Similar results of rboust network validation for OSPF routing scheme are shown \cref{sec:FERN-OSPF}, indicating the potential of FERN to benefit robust network design in real-world routing schemes.

\textbf{Network upgrade optimization}:
Compared to the robust validation problem, network upgrade optimization requires much more physical memory to build up the decision variables and constraints for the failure scenarios. 
In particular, we find that for a topology with more than 45 links, running the plain optimization problem will exceed
the 256 GB sever memory limit.
We show the time reduction and performance of FERN-based network upgrade algorithm in Fig. \ref{Fig:network-upgrade}, where the worst-case congestion is normalized by the results of plain optimization approach, and speed up ratio is obtained by dividing the time overhead of plain optimization problem by the time overhead of FERN-based algorithm. 
In evaluation, we normalize the input traffic matrices to keep worst-case congestion $MLU(x^w, G, D, r^o)=1.25$ so that the solution of FERN-based algorithm should be optimal if all the critical failure scenarios are included in FERN generated critical failure set $X_C$.    
We can find that the FERN-based network upgrade algorithm achieves exact optimal performance on most topologies by only considering critical failures.
In addition, FERN provides more than 10x time reduction over most topologies, and in some cases, the time reduction can be more than 100x.
Moreover, we can also expect a larger speed up ratio and a fully optimal solution with the help of P2 traing.
Further, we note that the FERN-based algorithm requires much less memory and could calculate the solution for the topologies with up to 80 links, which is not shown in Fig. \ref{Fig:network-upgrade}.

\textbf{Fault-tolerant traffic engineering:} 
Similar to network upgrade optimization, the memory use of the plain fault tolerant traffic engineering problem in Eq.\ref{eq:R3-origin} is also proportional to the number of failure scenarios.
In particular, we find that for a topology with more than 70 links, running the plain optimization problem will exceed
the 256 GB sever memory limit.
We show the time reduction and performance of the FERN-based fault tolerant TE algorithm in Fig. \ref{Fig:network-upgrade}.
We can find that the FERN-based fault tolerant TE algorithm provides comparable performance on almost all the topologies, and the solution is also validated to be congestion-free after the extra iteration step. 
In addition, the FERN-based algorithm requires less time overhead for most test cases and achieves up to 9x speed up for some middle-scale topologies, indicating good scalability.
Further, we note that the FERN-based algorithm requires much less memory and could calculate the solution for all the middle-scale topologies, which have up to 107 links.

\begin{table*}[htbp]\footnotesize
    \caption{Speed up ratio of FERN-based algorithm on robust network design problems. The network scale is measured with the number of links. Minimum, maximum and average speed up ratio from left to right. Note that $\infty$  represents that the plain optimization problem is infeasible to solve.}
    \label{tab:FERN-scalability}
    \centering
    \scalebox{0.9}{
    \begin{tabular}{c|c|ccc}
        \toprule
        \multirow{2}*{\shortstack{\textbf{Network} \\ \textbf{Scale}}} & \multicolumn{1}{|c}{\textbf{BRITE}} & \multicolumn{3}{|c}{\textbf{Topology Zoo}}  \\
        \cline{2-5}
        &  \textbf{[0,30]}  &  \textbf{[0, 30)} & \textbf{[30, 60)} & \textbf{[60, 110)} \\
        \midrule
         Robust Validation & [6.81, 85.7], 25.5  & [2.98, 42.6], 8.21 & [2.75, 83.4], 22.8 & [6.65, 61.9], 24.5 \\
        Network Upgrade & [3.72, 202], 56.9 & [1.52, 61.2], 12.0 &  [2.94, 173], 36.2 & $\infty$  \\
        Fault Tolerant TE & [0.549, 9.04], 2.96 & [0.419, 6.88], 1.30  & [0.314, 10.3], 2.75 & [1.31, 7.82], 4.09  \\
        \bottomrule
        \end{tabular}}
\end{table*}

We show the speed up ratio of FERN-based robust network design algorithm compared to the plain optimization problem in Table \ref{tab:FERN-scalability}.
We can find that for all the three robust network design problems, the speed up ratio of FERN-based algorithms keeps increasing with the network scale.
Moreover, FERN-based algorithm achieves better speed up ratio over BRITE generated topologies compared to the topologies with similar scale in topology zoo.
We note that FERN better predict the failure impact of BRITE generated topologies, i.e., select less non-critical failure scenarios to form the predicted critical failure set.
We also note that less accuracy of general model on unseen large topologies causes the performance degradation of robust validation which is not shown on the table. 
We further evaluate the performance of FERN-based robust network design algorithms with ground-truth critical failure scenarios to show the potential of FERN after P2 training.
We find that with an ideal 100\% precise FERN model, FERN-based algorithms could provide up to 3600x speed up ratio for network upgrade optimization problem.
We show the additional evaluation results in \cref{sec:extra-results}. 

At the end of evaluation, we note that FERN is until now not a perfect approach achieving optimal solution and best speed up ratio to all the testing cases.
But the evaluation results have shown the great potential of FERN to solve the common kernel of robust network problems and enhance the scalability of existing robust network design approaches.

\section{Related work}
\textbf{Network robust validation}:
\cite{chang2017robust} formalizes the robust validation problem as a two-stage max-min optimization problem and then reduces the computational overhead of solving the original two-stage optimization by relaxation skills. 
Netdice \cite{steffen2020probabilistic} improves the computational overhead of network availability validation by cut-off some non-critical link failures in the network.
However, such techniques of speeding up the network robust validation are only suitable for limited application scenarios and are still not enough for large-scale networks.

\textbf{Robust network planning}:
\cite{ahuja2021capacity} introduces the network planning in Facebook WAN, where 500 failure scenarios are protected based on historical data, including 300 single-fiber failures and 200 multi-fiber failures.
NeuroPlan \cite{zhu2021network} uses GNN to model the network structure information and combine Graph Neural Network (GNN) and Reinforcement Learning (RL) to prune the search space for the network planning optimization problem.
{\rev
 It is important to note that our work is orthogonal to NeuroPlan. While NeuroPlan prunes the solution space of the network planning problem by generating network planning solutions (i.e., link capacities) using a trained GNN model and then solving the Integer Linear Programming (ILP) problem with maximum capacity constraints based on the GNN-generated plan, FERN focuses on reducing the constraints for reliability under various failure scenarios. We also note that FERN can potentially enhance the training process of NeuroPlan by swiftly validating whether the generated plan satisfies traffic demand under various failure scenarios. 
 Furthermore, existing approaches for network planning typically select a set of failure scenarios for protection, and FERN could assist in further pruning the problem scale, thereby improving scalability.
 }

\textbf{Fault-tolerant traffic engineering}:
{\rev
Modeling link failures poses a significant challenge in traffic engineering problems. FFC \cite{liu2014traffic} introduces a robust traffic engineering scheme based on a tunnel-based TE approach. However, PCF \cite{jiang2020pcf} identifies performance degradation in FFC and resolves the issue through a local proportional routing scheme and a new optimization model. TEAVAR \cite{bogle2019teavar} optimizes network performance with a probabilistic guarantee. Nevertheless, the huge number of failure combinations makes it challenging to design a scalable algorithm with optimal performance for existing fault-tolerant traffic engineering problems.
R3 \cite{wang2010r3} provides a link failure resilient routing scheme based on link backup reroute.
\cite{chang2019lancet} shows that R3 could not work where no congestion-free routing solution exists to protect all the failure cases and propose an algorithm, Lancet, to protect as many failure cases as possible.
We note that R3 and Lancet employ duality and relaxation techniques to reduce the exponential number of constraints in the linear programming (LP) optimization problem by combining a class of failure scenarios (e.g., $f$ simultaneous failures or a specific subset) into a single constraint representation.
However, these duality and relaxation techniques used for link backup cannot be directly applied to other formulations of robust network design problems, such as optimal centralized rerouting and robust network validation. Moreover, FERN has the potential to assist Lancet by rapidly evaluating failure scenarios that cannot be protected. 
}
Compared to existing approaches for fault-tolerant traffic engineering, FERN proposes a novelty principal to design an efficient fault-tolerant traffic engineering algorithm focusing on protecting a small subset of failure scenarios.

\textbf{GNN for network modeling and optimization}:
{\rev 
RouteNet-Erlang \cite{ferriol2022routenet} uses GNN to model the network performance (e.g., flow latency, throughput, packet loss and queueing delay).
Such an ML-based network QoS model act as an efficient network simulator for evaluating and fine-tuning routing decisions \cite{rusek2020routenet}.
We note that FERN models a more complicated performance metric, namely failure impact, which requires a comprehensive understanding of the input network state at both local and global levels.
Additionally, there are existing studies that optimize network performance using GNNs. For instance, \cite{mao2019learning} applies GNN to model job relationships and employs a reinforcement learning (RL) algorithm to optimize job scheduling for incoming network application requests. Similarly, \cite{bernardez2021machine} adopts a GNN model to abstract the network state and employs RL to address the traffic engineering problem by minimizing network congestion.
However, such algorithms are trained for specific tasks and lack a generalization guarantee for unseen application scenarios (e.g., unseen large topologies).
Furthermore, it is worth noting that most existing ML-based traffic engineering solutions either overlook failure resiliency or only react to identified failure scenarios passively. 
Such passive network decision-making schemes may cause the network to fail to respond to failures in a timely manner.
}


\section{Conclusion}
In this paper, we propose FERN, applying graph attention networks to enhance the failure impact evaluation and robust network design. 
First, we show that evaluating the impact of a huge number of failure scenarios is a common kernel problem to robust network design, which brings huge computational overhead and limits scalability.
We next propose a GAT-based algorithm to predict failure impact and select critical failure scenarios efficiently.
Meanwhile, we recast three typical robust network design problems using the failure impact evaluation function in this paper.
We enhance the tractability and scalability of the recast robust network design problems with the predicted critical failure scenarios.
We evaluate FERN over hundreds of real-world network topologies for failure evaluation and fault tolerant network design.
The experiment results validate that FERN efficiently and accurately predicts the impact of given failure scenarios and generalizes to unseen network topologies.
Extensive experiments show that FERN significantly improves the tractability and scalability of existing robust network design problems with negligible performance gap.

\bibliographystyle{IEEEtran}
\bibliography{sample-base}

\clearpage
\appendices

\section{Notations}
We show the notations in this paper in Table. \ref{tab:my_label}.
\begin{table}[htbp]\footnotesize
    \centering
    \begin{tabularx}{250pt}{|l|X|}
    \hline
    \textbf{Symbol}  & \textbf{Description} \\
    \hline
    $G=(V, E)$  & Network topology.\\
    $D$  & Set of traffic demands.\\
    $T(d, v)$ & Traffic of demand $d$ at node $v$, positive at source, negative at destination and 0 otherwise.\\ 
    $C_e$ & Capacity of link $e$. \\
    $r$ & Route decisions. \\ 
    $r^o$ & Optimal route decisions. \\
    $\delta$ & Congestion level threshold. \\
    $MLU$ & Maximum link utilization.\\
    $\phi$ & Scenario with no link failure.\\
    $X_f$  & Set of $f$ simultaneous link failure scenarios.\\
    $X_C$  & Set of critical failure scenarios.\\
    $a_e$  & Augmented capacities to a link. \\
    $x^w$  & Failure scenario with largest failure impact. \\
    \hline
    \end{tabularx}
    \caption{Notations in this work.}
    \label{tab:my_label}
\end{table}

\section{Proof for theorem 1}
\label{sec:proof-network-upgrade}
\begin{proof}
Clearly, the augmented topology $G+\Delta G$ will never increase the congestion level.
That is, for any $x\in X$, we have
\begin{equation}
    MLU(x, G+\Delta G, D, r^o) \leq  MLU(x, G, D, r^o).\notag
\end{equation}
Combined with Eq. \eqref{eq:impact-def-FERN}, we have
\begin{equation}
\begin{aligned}
        F_\theta(x, G+\Delta G, D, r^o) \cdot MLU(\phi, G+\Delta G, D, r^o) \\
        \leq F_\theta(x, G, D, r^o) \cdot MLU(\phi, G, D, r^o).
\end{aligned}
    \label{eq:proof-1}
\end{equation}
According to Eq. \eqref{eq:network-upgrade-F} and \eqref{eq:proof-1} we have, if the failure scenario $x$ satisfies
\begin{equation}
    F_\theta(x, G, D, r^o) \cdot MLU(\phi, G, D, r^o) \leq 1,
    \label{eq:proof-2}
\end{equation}
$x$ also satisfies Eq. \eqref{eq:network-upgrade-F}.
Thus only the critical failure scenarios causing network congestion to original topology $G$ need to be validated in network upgrade optimization.
Since the congestion constraint in Eq. \eqref{eq:proof-2} is also equivalent to the constraint in Theorem \ref{theory:network-upgrade} combined with Eq. \eqref{eq:impact-def-FERN}, thus showing the result as in the statement of the Theorem.
\end{proof}

\section{Fault-tolerant traffic engineering algorithm design}
\label{sec:Robust-TE-appendix}
We constrain $U'_\mathcal{E}$ for single link failure $x^\mathcal{E} \in X_1$ as
\begin{equation}
    \sum_{d} r(d, e) + \sum_{l} \hat{r}(l, e) x^\mathcal{E}_l  \leq U'_{\mathcal{E}} C_e, \forall e \in E
\end{equation}
and only the performance of critical failure scenarios are further considered
\begin{equation}
     \sum_{d} r(d, e) + \sum_{l} \hat{r}(l, e) x_l  \leq U_C C_e, \forall e \in E, x \in X_C
\end{equation}
We further constrain $U_C \leq 1$ to make the results congestion free under the critical failure scenarios.
The whole modified optimization problem is shown as follows
\begin{equation}
\begin{gathered}
    \min_{r,\hat{r}}~ U_C + \frac{1}{|E|}\sum_{\mathcal{E}} U'_\mathcal{E} \\
    s.t.\\
     \sum_{e_{src}=v}r(d, e) - \sum_{e_{dst}=v}r(d, e) = T(d, v), \forall v \in V, d\in D \\
     r(d, e) \geq 0, \forall d \in D, e \in E  \\
     \sum_{e_{src} =v}\hat{r}(l, e) - \sum_{e_{dst}=v}\hat{r}(l, e) = T(l, v), \forall v \in V, l \in E  \\
     \hat{r}(l, e) \geq 0,  \forall l\in E, e \in E  \\
       \sum_{d} r(d, e) + \sum_{l} \hat{r}(l, e) x_l  \leq U_C C_e, \forall e \in E, x \in X_C  \\
       \sum_{d} r(d, e) + \sum_{l} \hat{r}(l, e) x^\mathcal{E}_l  \leq U'_{\mathcal{E}} C_e, \forall e \in E, x^\mathcal{E} \in X_1\\
       U_C \leq 1
\end{gathered}
      \label{eq:R3-new}
\end{equation}

\section{Details of experiments}

\subsection{Model hyper-parameter}
\label{sec:hyper-parameter}
\begin{table}[h]
        \caption{Hyper-parameter setting.}
        \label{tab:model-setting}
        \centering
        \begin{tabular}{cc}
            \toprule
            \textbf{Hyper-parameter} & \textbf{Value}\\
            \midrule
            Input units & 16 \\
            Hidden units & 64 \\
            Global attention heads & 4 \\
            Local attention heads & 4 \\
            \bottomrule
            \end{tabular}
    \end{table}
We show the hyper-parameter in Table \ref{tab:model-setting}.

\subsection{Details of training}
\label{sec:training}
Training a large and general model over such a huge dataset is not an easy task.
We note that the classification problem in \cref{sec:loss} is easier than the regression problem.
In addition, we find that FERN model converges faster on small datasets.
So, we first train a classification model over 10\% data from \emph{BRITE-small-train} dataset to obtain a set of initial model parameters.
Then we train the classification model over the full dataset for about 1000 epochs to obtain the general models.
Finally, we initialize the regression model with the model parameters of the trained general classification model and train it for another 500 epochs. 
We note that such a pre-training process over a small dataset can make the model converge much faster, and the whole P1 training process takes about ten days on the server.

We note that in P2 training, the model uses the worst-case failure impact of selected failure scenarios rather than all failure scenarios to approximate the worse case failure impact, and such an approximation shows good performance for most scenarios.
In real world deployment, we also suggest that a collection of historical critical failure scenarios could benefit the P2 training. 


\subsection{FERN for OSPF}
\label{sec:FERN-OSPF}
\begin{table}[ht]\footnotesize
        \caption{Dataset information for the OSPF routing.}
        \label{tab:dataset-info-OSPF}
        \centering
        \scalebox{0.8}{
        \begin{tabular}{cccc}
            \toprule
            Dataset & Topology number & Topology scale & TM number\\
            \midrule
            BRITE-small-train & 90 & 6-15 & 1000\\
            Topology-zoo-small-train & 69 & $<$ 20 & 1000 \\
            Topology-zoo-middle-train & 78 & 20 - 80 & 100 \\
            \midrule
            BRITE-small-test1 & 90 & 6-15 & 100 \\
            Topology-zoo-small-test1 & 69 & $<$ 20 & 100 \\
            Topology-zoo-middle-test1 & 78 & 20 - 80 & 100 \\
            \midrule
            BRITE-small-test2 & 10 & 6-15 & 100 \\
            Topology-zoo-small-test2 & 10 & $<$ 20 & 100 \\
            Topology-zoo-middle-test2 & 20 & 20 - 80 &100 \\
            \midrule
            Topology-zoo-large-test & 11 & 80 - 166 & 100 \\
            \bottomrule
            \end{tabular}}
    \end{table}
The information of the dataset for the OSPF routing scheme is shown in Table \ref{tab:dataset-info-OSPF}.
We train a general model for OSPF routing (denoted as FERN-OSPF) with a similar P1 training process with the optimal MCF routing.

\begin{figure*}[htbp]
\centering
\begin{minipage}[t]{0.3\textwidth}
\centering
\includegraphics[width=\linewidth]{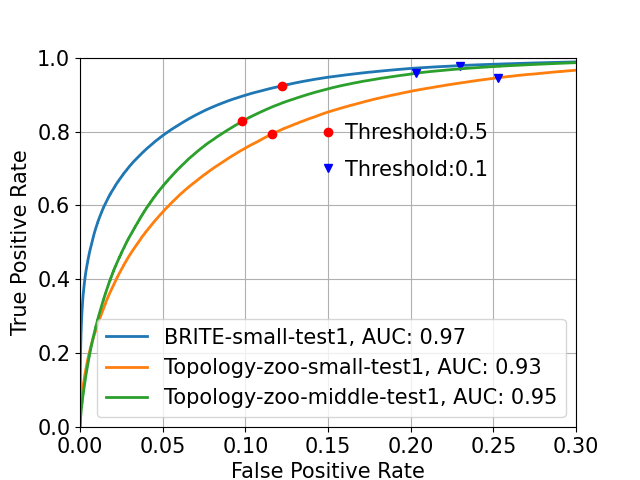}
\captionsetup{font={footnotesize}}
\caption{\footnotesize Accuracy (ROC curves) of FERN-OSPF classification model on seen topologies.}
\label{Fig:P1-ROC-test1-OSPF}
\end{minipage}
\hspace{5mm}
\begin{minipage}[t]{0.3\textwidth}
\centering
\includegraphics[width=\linewidth]{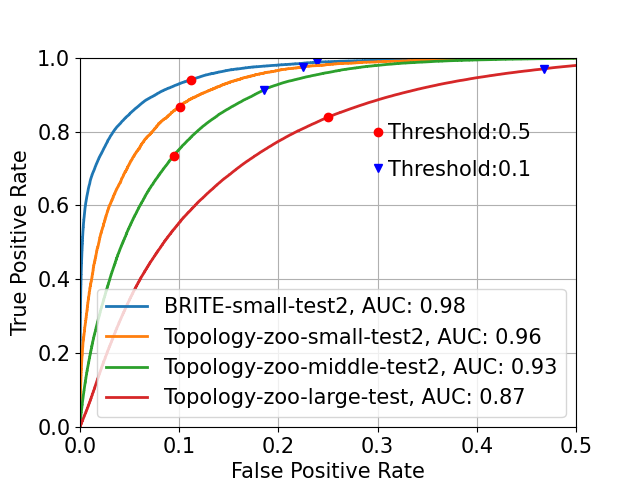}
\captionsetup{font={footnotesize}}
\caption{\footnotesize Accuracy (ROC curves) of FERN-OSPF classification model on unseen topologies.}
\label{Fig:P1-ROC-test2-OSPF}
\end{minipage}
\hspace{5mm}
\begin{minipage}[t]{0.3\textwidth}
\centering
 \includegraphics[width=\linewidth]{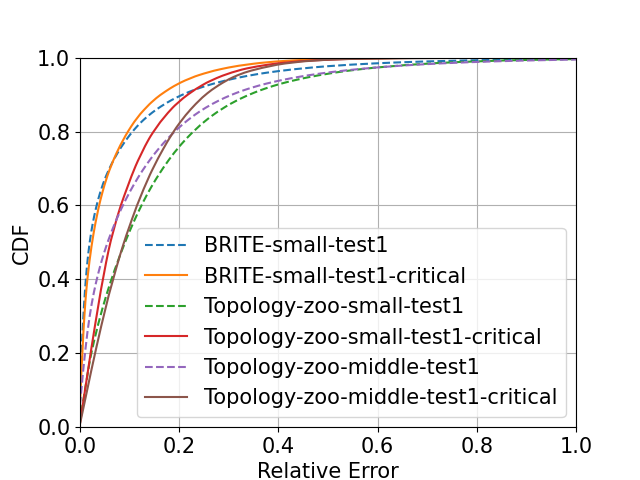}
\captionsetup{font={footnotesize}}
\caption{\footnotesize Relative error of FERN-OSPF regression model on seen topologies.}
\label{Fig:P1-mre-test1-OSPF}
\end{minipage}
\hspace{5mm}
\begin{minipage}[t]{0.3\textwidth}
\centering
 \includegraphics[width=\linewidth]{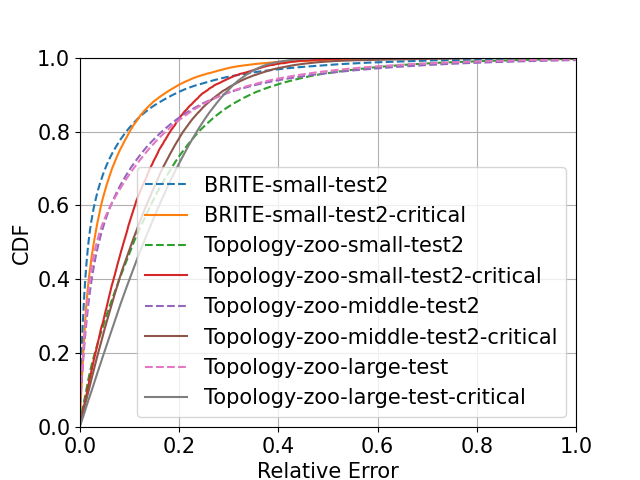}
\captionsetup{font={footnotesize}}
\caption{\footnotesize Relative error of FERN-OSPF regression model on unseen topologies.}
\label{Fig:P1-mre-test2-OSPF}
\end{minipage}
\begin{minipage}[t]{0.3\textwidth}
\centering
\includegraphics[width=\linewidth]{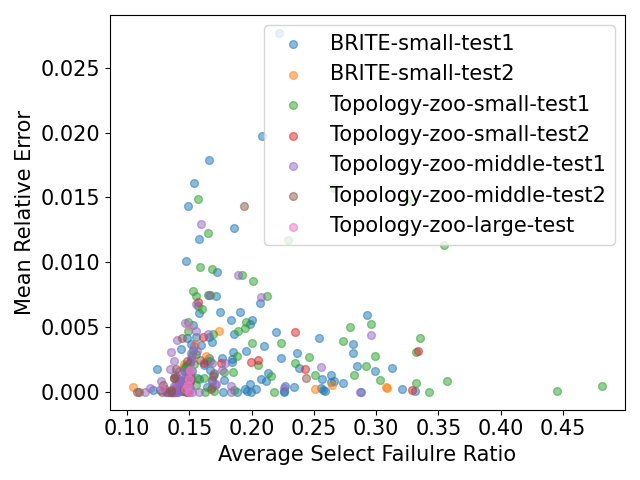}
\captionsetup{font={footnotesize}}
\caption{\footnotesize Performance of FERN-based robust validation for OSPF routing.}
\label{Fig:robust-validation-OSPF}
\end{minipage}
\begin{minipage}[t]{0.3\textwidth}
\centering
\includegraphics[width=\linewidth]{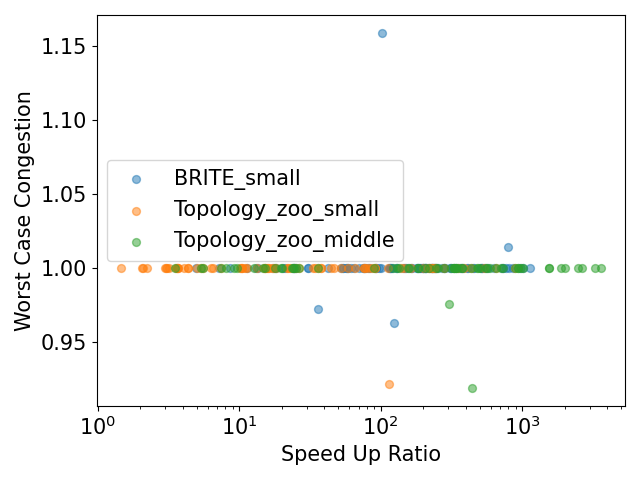}
\captionsetup{font={footnotesize}}
\caption{\footnotesize Performance of applying FERN with ground-truth output in network upgrade optimization.}
\label{Fig:network-upgrade-ground-truth}
\end{minipage}
\hspace{5mm}
\begin{minipage}[t]{0.3\textwidth}
\centering
\includegraphics[width=\linewidth]{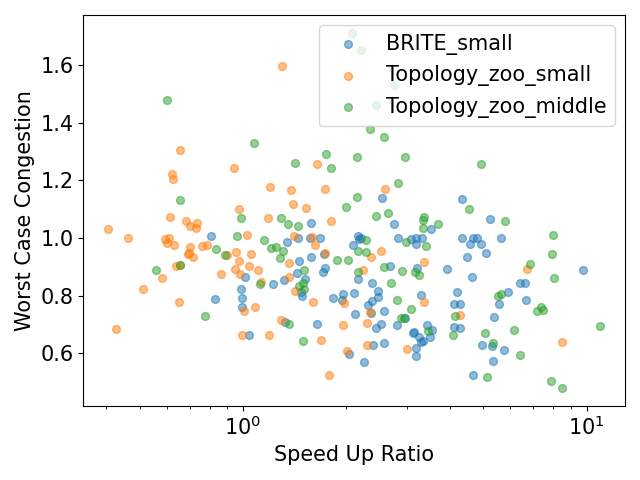}
\captionsetup{font={footnotesize}}
\caption{\footnotesize Performance of applying FERN with ground-truth output in fault tolerant traffic engineering.}
\label{Fig:robust-TE-ground-truth}
\end{minipage}
\end{figure*}

The results of failure evaluation for FERN-OSPF are shown in Fig. \ref{Fig:P1-ROC-test1-OSPF}, Fig. \ref{Fig:P1-ROC-test2-OSPF}, Fig. \ref{Fig:P1-mre-test1-OSPF}, and Fig. \ref{Fig:P1-mre-test2-OSPF}.
We note that these results are similar to FERN-MCF, indicating a good generalizability of our proposed GAT-based failure evaluation function to routing scheme that are difficult to express with LP models.

We also show the effectiveness of FERN-OSPF in speeding up robust network validation for OSPF routing.
The results are shown in Fig. \ref{Fig:robust-validation-OSPF}.
It validates that FERN-OSPF could speed up robust network validation for OSPF routing with negligible worst case performance prediction error over topologies of different scales.

\subsection{Ideal experiment results}
\label{sec:extra-results}

We further evaluate the ideal performance of FERN with ground-truth outputs, i.e., assume FERN figure out the critical failure scenarios with a 100\% accuracy.
The results are shown in Fig. \ref{Fig:robust-TE-ground-truth}.
We can find that with a 100\% accurate critical failure prediction, the speed up ratio can be up to 3600x for medium-sized topologies.
Meanwhile, the optimal performance guarantee in Theorem \ref{theory:network-upgrade} also holds.
Moreover, with a ideal FERN model with ground-truth outputs, we could obtain the solution for the robust validation problem directly.
We note that with P2 training we could obtain a much more precise FERN model, which provides close performance to the ideal FERN model.
The results above also show the great potential of FERN to further enhance the scalability of robust network design problem in the future.

\end{document}